\documentclass[a4paper,11pt]{article}
\pdfoutput=1 

\usepackage{jcappub} 

\usepackage{aasmacros}
\usepackage[T1]{fontenc} 
\usepackage{cprotect}
\usepackage{multirow}
\usepackage{diagbox}
\usepackage{floatrow}

\newfloatcommand{capbtabbox}{table}[][\FBwidth]

\title{\boldmath Comparison of delensing methodologies and assessment of the delensing capabilities of future experiments}

\author[a,b]{P. Diego-Palazuelos,}
\author[a]{P. Vielva,}
\author[a]{E. Mart\'inez-Gonz\'alez}
\author[a]{and R. B. Barreiro}
\affiliation[a]{Instituto de F\'isica de Cantabria (CSIC-Universidad de Cantabria),\\ 
Avda. de los Castros s/n, E-39005 Santander, Spain}
\affiliation[b]{Dpto. de F\'isica Moderna, Universidad de Cantabria, \\
Avda. los Castros s/n, E-39005 Santander, Spain}

\emailAdd{diegop@ifca.unican.es}
\emailAdd{vielva@ifca.unican.es}
\emailAdd{martinez@ifca.unican.es}
\emailAdd{barreiro@ifca.unican.es}

\abstract{Most of the CMB experiments proposed for the next generation aim to detect the Primordial Gravitational Wave Background (PGWB). The fulfillment of this objective depends on our capacity to separate Galactic foreground emissions and to \emph{delens} the secondary B-mode component induced by weak gravitational lensing. Focusing on the latter of these efforts, in this work we briefly review the basic aspects of lensing, and exhaustively compare the performance of current delensing methodologies and implementations within the Born approximation as a preparation for the analysis of the data to come in the following years. Two of the main conclusions that can be drawn from our study are that, for next-generation experiments, delensing efficiency will still be limited by the quality of the data itself rather than by the limitations of current delensing methodologies, and that template delensing within the antilensing approximation will be the optimal (balancing accuracy and computational cost) technique to employ. We then evaluate the delensing capabilities of future experiments (like the Simons Observatory, the CMB Stage-IV, or the LiteBIRD and PICO satellites) by applying that methodology onto numerical simulations of the typical CMB and lensing potential reconstructions that they are expected to produce, and quantify how internal and external delensing will help them to improve their sensitivity to detect the PGWB. We also consider the benefits that a joint analysis of their data would provide.}

\begin{document}
\maketitle
\flushbottom

\section{Introduction}
\label{sec:intro}

The existence of a stochastic Primordial Gravitational Wave Background (PGWB), formed when microscopic quantum fluctuations of the metric were stretched up to super-horizon scales by the sudden expansion of space-time that occurred during inflation~\cite{caprini}, is a common prediction in the majority of inflationary models. If detected, the PGWB would not only prove that indeed the universe underwent an inflationary period, but would also provide a large amount of information about the physics of inflation~\cite{watanabe&komatsu}, ranging from the energy scale at which it took place to the type of field that drove it. Since it has been able to free-stream from times as early as (possibly) Planck scales, the PGWB also has the potential of becoming one of the most powerful cosmological probes, donning information about the phase transitions and particle creation/annihilation that took place in the early universe, and allowing new independent measurements of cosmological parameters.\\

Unfortunately, according to the current constraints that CMB measurements impose on inflation~\cite{planck2018inflacion} (see below), the amplitude of the PGWB is expected to be too small to allow a direct detection with any of the present-day or near-future gravitational wave interferometers~\cite{GW_detection_w/_interferometry} Nevertheless, given the tensor-like nature of the metric perturbations that conform it, the PGWB also has the property of polarizing photons~\cite{polnarev, zhao, cabella&kamionkowski}, introducing an additional B-mode component to the CMB polarization. Current measurements constrain the amplitude of this primordial B-mode component (controlled by the ratio between tensor and scalar perturbations $r=\mathcal{P}_t(k)/\mathcal{P}_s(k)$) to be $r<0.056$~\cite{planck2018inflacion}, and is up to future experiments to bring the detection threshold down to the $r\sim 0.001$ values which would allow the detection of physically motivated inflationary models~\cite[e.g.,][]{grishchuk, starobinski}.\\

However, the faint signal that the PGWB leaves on the CMB polarization is vastly obscured by those of astrophysical (notably Galactic) foreground emissions and the secondary B-mode component induced by the weak gravitational lensing that CMB photons suffer while traversing the large-scale structure present in the late universe. As CMB experiments continue to lower their sensitivities and the performance of component separation techniques improves, the development of a methodology to revert the effects of lensing, commonly known as \emph{delensing}, become very important if we want to succeed at our goal of PWGB detection. Like \cite{first_delensing_w/_CIB} first demonstrated on data for temperature, the effects of lensing can be reverted if the matter distribution causing the deflection of photons is known. In that work, the Cosmic Infrared Background (CIB), an unresolved emission coming from dusty star-forming galaxies at similar redshifts than the matter distribution causing the lensing, was used as a proxy for the lensing potential. Shortly after, \cite{planck_first_internal_delensing_temperature&polarization} extended the formalism applied by \cite{first_delensing_w/_CIB} to the CMB polarization, using this time a reconstruction of the lensing potential coming from the CMB itself. In more recent works, more sophisticated reconstructions of the lensing potential have been successfully applied to the delensing of polarization data from the Planck satellite~\cite{planck2018lensing}, and ground-based experiments like SPTPol~\cite{delensing_sptpol_w/_herschel} and POLARBEAR~\cite{template_delensing_polarbear}, and delensing techniques have shifted from the antilensing used in earlier works to template delensing.\\

In preparation for the intensive work to come in the following years, we wanted this document to offer an overview of the current and future panorama of the delensing field. With this purpose, we dedicate the first half of this document to collect and compare the performance of the various lensing implementations devised in the last fifteen years. We limit ourselves to implementations within the Born approximation, excluding therefore all ray-tracing algorithms. We also review several delensing methodologies, and, via numerical simulations, compare the delensing efficiency they can achieve when implemented with the different lensing algorithms. This exercise not only has the objective of finding the optimal technique, but also determining if current tools would be enough to fully exploit the data to come in the next decade. Although partial comparisons of lensing codes have been previously made \cite{lens2hat, flints, taylens, lenseflow}, a complete and coherent comparison like the one here presented of, both, implementations and methodologies, was lacking in the literature.\\

To address the future of the field, we dedicate the second half of this work to predict the delensing capabilities of future experiments like the Simons Observatory, the CMB Stage-IV, or the LiteBIRD and PICO satellites. Once again, we will base our predictions on numerical simulations of both, the typical CMB maps, and potential reconstructions that those experiments are expected to produce, extending some of the previous analytical forecasts done in the literature \cite{delensing_w/_external_datasets, errard, simons_observatory, s4, pico}. In our predictions, we take special care to always specify the quality of the potential reconstruction that is being used. This piece of information is often omitted in the literature, impeding the comparison of results between different works.\\

This work is structured as follows. Section~\ref{sec:physics_of_lensing} offers a quick review on the physics of lensing. In section~\ref{sec:lensing_algorithms} we collect and review the various lensing algorithms devised in the last fifteen years, leaving the comparison of their performance and application to different delensing methodologies to section~\ref{sec:delensing_methodologies_general}. Having selected the optimal delensing methodology for next-generation experiments, in section \ref{sec:predicciones_generales} we will then determine how well can B-mode polarization be delensed, as a function of the properties of the input CMB map and the lensing potential reconstruction. The contribution that delensing would offer to the detection of the PGWB in the context of future CMB experiments will then be quantified in section~\ref{sec:detectability}. We leave our final conclusions and discussion on some of the remaining challenges that the delensing community still has to affront to section \ref{sec:conclusiones}.\\


\section{Physics of lensing}
\label{sec:physics_of_lensing}

In this section we offer a brief overview of the physics of lensing (if needed, a more thorough review can be found in~\cite{lewis&challinor_lensing_review}). In section~\ref{sec:weak_lensing} we recall the main effects that weak gravitational lensing induces on the CMB radiation. Also, given that a good estimate of the lensing potential is an essential requirement for delensing, a brief commentary about the kind of potential reconstructions that could be produced from future CMB experiments is made on section~\ref{sec:lensing_potential}.\\

\subsection{Weak gravitational lensing of the CMB}
\label{sec:weak_lensing}

On their way to us, CMB photons are subjected to the gravitational pull exerted by the large-scale structure present in the late universe. According to General Relativity, photons traveling near a mass distribution will experiment a transverse acceleration given by the gradient of the gravitational potential ($\Phi$), effectively deflecting the photons direction of motion. Adding up all the small local deflections caused from the matter distribution CMB photons encounter in their journey from the last scattering surface (at comoving distance $\chi^*$) to us, we get a total deflection angle of~\cite{lewis&challinor_lensing_review}
\begin{equation}\label{eq:total_deflection_angle}
    \vec{\alpha}(\vec{n})=-2\int^{\chi^*}_0d\chi\cfrac{f_K(\chi^*-\chi)}{f_K(\chi^*)f_K(\chi)}\vec{\nabla}\Phi(\chi\vec{n}, \eta_0-\chi),
\end{equation}
where the $\vec{\nabla}$ operator denotes covariant derivative on the sphere, and $f_K(\chi)$ is the \emph{angular diameter distance}, i.e., the function that relates comoving distances to subtended angles on the sky depending on the curvature of space. The average deflection angle that CMB photons suffer is of about $\sim 2$ arcmin, justifying the validity of the \emph{weak lensing approximation} implicitly used in the derivation of equation \eqref{eq:total_deflection_angle}.\\

Lensing by transverse gradients does not change the frequency distribution of photons, ensuring that the lensed CMB has the same blackbody spectrum than the unlensed one. The number of photons per solid angle is also unaffected since lensing conserves surface brightness, because although the lensing induced magnification does increase the number of photons received from a certain patch, the angular size subtended by that patch also increases proportionally. The only apparent effect lensing seems to have then is to move photons around; an effect that would otherwise pass unnoticed for a perfectly isotropic CMB, and that is only appreciable because of the CMB anisotropies.\\

\begin{figure}[tbp]
\centering 
\includegraphics[width=0.9\textwidth]{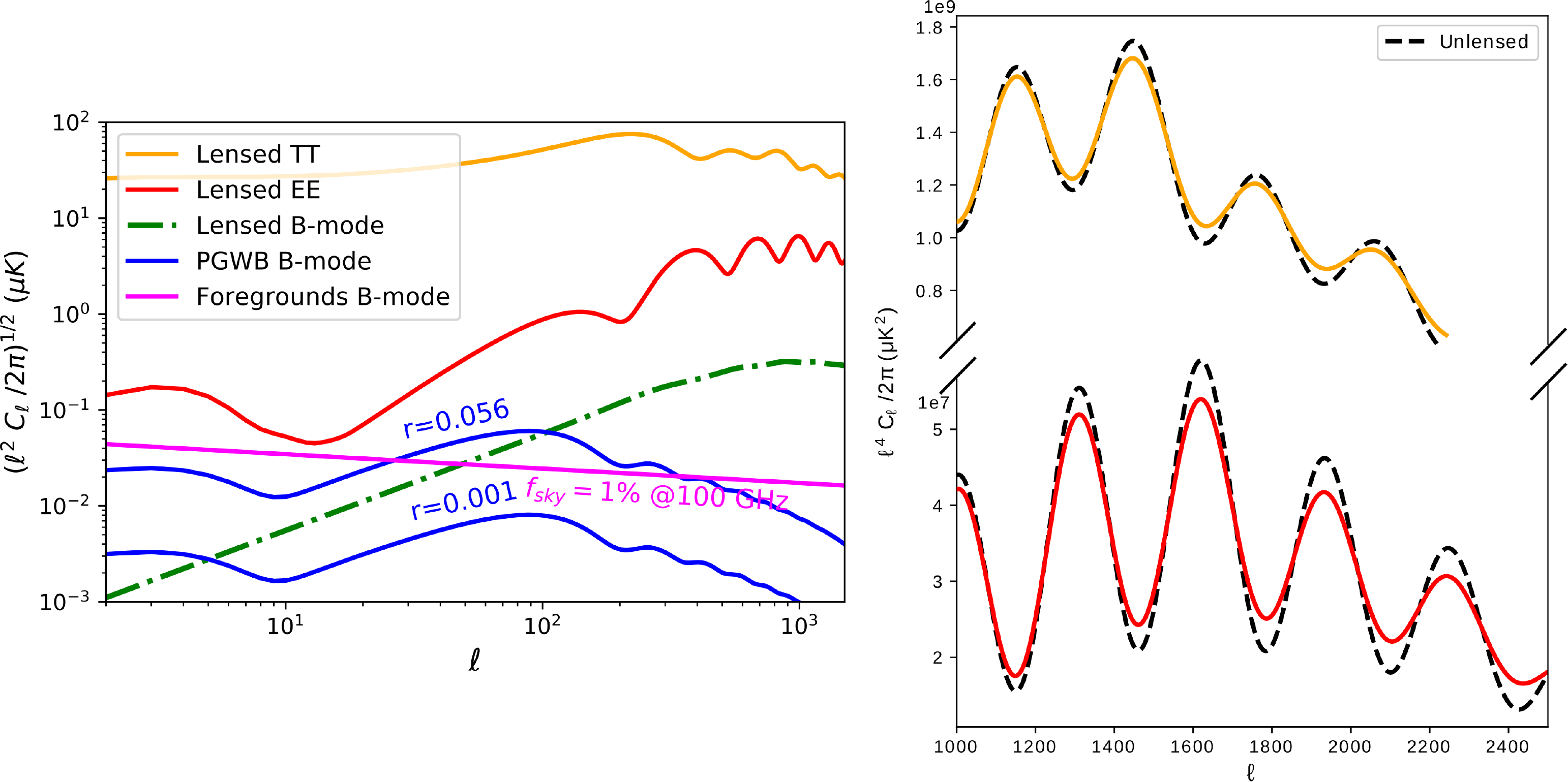}
\hfill
\caption{\label{fig:lensing_dampening_of_acoustic_oscillations} Temperature and polarization angular power spectra of the lensed CMB. Two amplitudes of the PGWB induced B-mode polarization are included: one to show the current $r<0.056$ upper limit set by the Planck mission~\cite{planck2018inflacion}, and another one to reflect the $r=0.001$ sensitivity level targeted by next-generation CMB experiments~\cite{simons_observatory, s4, pico}. Note that in the closer view shown on the right, $C_\ell$s are multiplied by a $\ell^4$ factor to highlight the subtle dampening of the acoustic oscillations that lensing produces. Although not shown here, the $TE$ cross-spectrum is also smoothed accordingly. For future reference, the foreground B-mode spectrum expected at the $1\%$ cleanest fraction of the sky at 100 GHz~\cite{errard} is also included (approximately a $F_\ell=1.5\times 10^{-2}\ell^{-2.29}$ law).}
\end{figure}

After being lensed, the hot and cold spots of the anisotropy pattern would appear larger or smaller by $\sim 2$ arcmin, which broadens the spot size distribution and leads to a dampening of the arcminute-scale acoustic peaks in the $TT$ and $EE$ angular power spectrum of the CMB (see inset of figure~\ref{fig:lensing_dampening_of_acoustic_oscillations}). Formally, this effect is understood as a convolution between the angular power spectra of CMB fields and the matter distribution~\cite{hu_lensing_harmonic_approach}:
\begin{eqnarray}
\tilde{C}_\ell^{TT}=\left(1-\ell^2R^\phi\right)C_\ell^{TT}+
\int\cfrac{d^2 l'}{(2\pi)^2}[(l-l')\cdot l']^2C_{|l-l'|}^{\phi\phi}C_{\ell'}^{TT}; \hspace{15mm}\\
\tilde{C}_\ell^{EE}=\left(1-\ell^2R^\phi\right)C_\ell^{EE}+
\int\cfrac{d^2 l'}{(2\pi)^2}[(l-l')\cdot l']^2C_{|l-l'|}^{\phi\phi}C_{l'}^{EE}\cos^22(\varphi_{l'}-\varphi_{l}),
\end{eqnarray}
where $l$ indicates a point in harmonic space $l=(\ell,m)$ and the $\cos^22(\varphi_{l'}-\varphi_l)$ term is the rotation of the $l'$ base in the direction of $l$.  The $TE$ spectrum (although not shown in figure~\ref{fig:lensing_dampening_of_acoustic_oscillations}) is also smoothed accordingly:
\begin{equation}
\tilde{C}_\ell^{TE}=\left(1-\ell^2R^\phi\right)C_\ell^{TE}+
\int\cfrac{d^2 l'}{(2\pi)^2}[(l-l')\cdot l']^2C_{|l-l'|}^{\phi\phi}C_{\ell'}^{TE}\cos 2\varphi_{l'}.
\end{equation}

The $R^{\phi}$ term on the r.h.s. of these equations stands for half the total deflection angle power
\begin{equation}
R^{\phi}=\frac{1}{2}\langle|\nabla\phi|\rangle^2=\frac{1}{4\pi}\int\frac{d\ell}{\ell}\ell^4C_\ell^{\phi\phi},
\end{equation}
and, given the little power of $\ell^4C_\ell^{\phi\phi}$, its contribution can be neglected in a first approximation.\\

In addition, the transverse accelerations that photons suffer as they traverse the matter distribution also have the effect of turning E-mode polarization patterns into B-mode ones~\cite{lewis&challinor_lensing_review, cabella&kamionkowski}, introducing a B-mode component to the otherwise intrinsically purely E-mode CMB polarization. In terms of the angular power spectrum, this means that a B-mode signal will be generated from the convolution of the E-mode polarization field with the matter distribution: 
\begin{equation}
\tilde{C}_\ell^{BB}=\int\cfrac{d^2 l'}{(2\pi)^2}[(l-l')\cdot l']^2C_{|l-l'|}^{\phi\phi}C_{l'}^{EE}\sin^22(\varphi_{l'}-\varphi_l).
\end{equation}
This secondary B-mode signal, which is approximately equivalent to a $5\mu K\cdot$arcmin white noise up to $\ell\sim300$ (see the green dot-dashed line in figure~\ref{fig:lensing_dampening_of_acoustic_oscillations}), has already been detected by several of the currently operating ground-based CMB experiments~\cite{sptpol_lensed_B-modes_detection, actpol_lensed_B-modes_detection, polarbear_lensed_B-modes_detection, bicep/keck_lensed_B-modes_detection}. What is more, as instrument sensitivity and component separation continue to improve, these lensed B-modes will become a serious obstacle preventing the detection of the B-mode polarization induced by the PGWB for the next generation of experiments.\\

Finally, lensing is also a source of non-Gaussianity~\cite{lewis&challinor_lensing_review}, since, although it is a linear operation, its dependence with the gravitational potential is non-linear. In particular, the lensed CMB has both a non-zero bispectrum (or three-point correlation function), and a non-zero trispectrum (or four-point correlation function). In the next subsection, we will see how these higher order statistics can be used to recover a projected view of the matter distribution in the universe.\\

\subsection{Projected mass distribution across the sky: the lensing potential}
\label{sec:lensing_potential}

Equation \eqref{eq:total_deflection_angle} evidences how, apart from the curvature of space-time, the only other element necessary to determine lensing deflection angles is the gravitational potential describing how matter is distributed in the universe. Taking the spatial derivative out of the integral in \eqref{eq:total_deflection_angle}, the \emph{lensing potential} is then defined as
\begin{equation}
    \phi(\vec{n})\equiv-2\int^{\chi^*}_0d\chi\cfrac{f_K(\chi^*-\chi)}{f_K(\chi^*)f_K(\chi)}\Phi(\chi\vec{n}, \eta_0-\chi),
\end{equation}
\emph{i.e.}, the projection onto the sphere of the integrated mass distribution along the line-of-sight between us and the last scattering surface. In this way, if recombination is approximated to be an instantaneous process so that the CMB is emitted in a single source plane at $\chi=\chi^*$, and the very small effects of late-time sources and reionization are neglected, then all the information requiered for lensing is contained in a single two-dimensional map.\\ 

Deflection angles are therefore calculated from the lensing potential like $\vec{\alpha}(\vec{n})=\vec{\nabla}\phi(\vec{n})$. Once deflection angles are known, it should be possible to revert the effect of lensing just by remapping each point of the observed lensed CMB to its original position. This process would ideally allow us to recover the unlensed CMB, or at least to reduce the lensed B-mode component enough to facilitate a detection of the PGWB. Hence, although we would center our study around the \emph{delensing} process itself (dedicating the following sections to compare delensing methodologies and determine how the properties of the input CMB and lensing potential maps condition the attainable degree of delensing), obtaining a good estimate of the lensing potential is a crucial previous step for PWGB detection.\\

The lensing potential can be reconstructed from tracers of the large-scale structure of the universe~\cite{manzotti, multitracer} (like galaxy surveys~\cite{delensing_w/_ska}, the Cosmic Infrared Background~\cite{delensing_w/_cib, first_delensing_w/_CIB, delensing_sptpol_w/_herschel, planck2018lensing}, or tomographic line intensity mapping~\cite{line-intensity_mapping, delensing_w/_line-intensity_mapping}), or directly from the lensed CMB itself through the higher order statistics lensing introduces. Although known to be suboptimal at the low-noise regime expected for next-generation polarization maps~\cite{qe_are_suboptimal, hirata&seljak, hirata&seljak_eb_qe, lensit, qe_suboptimal_for_temperature, delensing_w/_external_datasets}, quadratic estimators~\cite{qe_formulas} are currently the most extended tool to obtain these \emph{internal} lensing potential reconstructions, and can be used to forecast what kind of reconstructions might be expected from future experiments since they provide a lower limit for the optimal \emph{maximum a posteriori}\footnote{Maximum a posteriori estimation is a take on Bayesian inference in which a single point estimate is returned instead of the full posterior probability distribution. As its name suggests, the MAP estimate chooses the point of maximal posterior probability. Although conceptually very similar to maximum likelihood estimation, MAP estimates have the advantage of allowing the prior to influence the choice of the point estimate.} (MAP) reconstructions~\cite{lensit, lenseflow,likelihood_reparameterization}. Given that lensed B-modes act as a noise for lensing potential reconstruction, better estimates can be achieved by repeatedly applying either quadratic or MAP estimators. In each step of this iterative process, the current estimate of the potential is used to partially remove the lensed B-modes, successively improving the resulting reconstruction.\\ 

\begin{figure}[tbp]
\centering 
\includegraphics[width=0.55\textwidth]{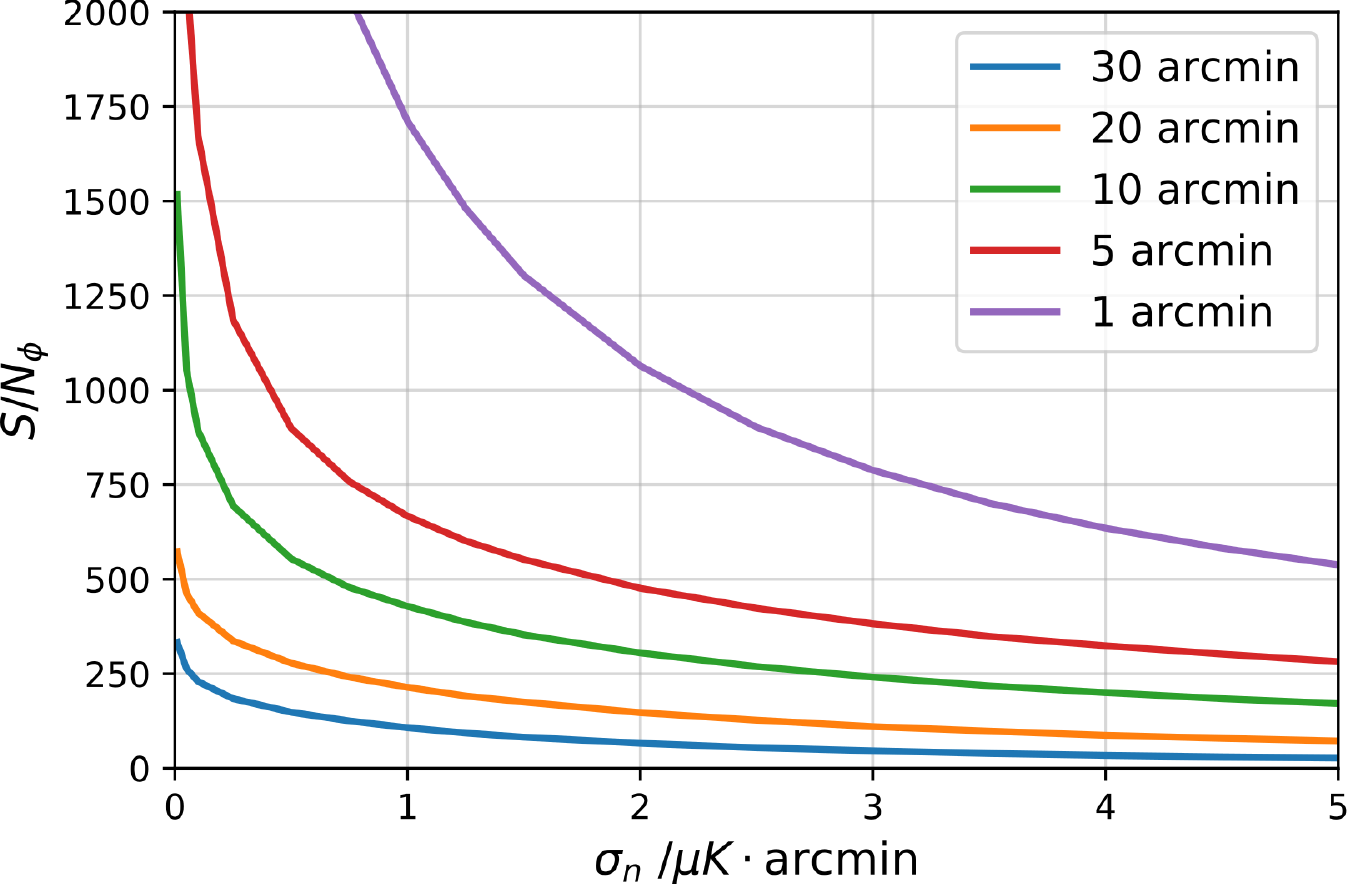}
\hfill
\cprotect\caption{\label{fig:sigmas_of_the_lensing_potential_reconstruction} Forecast on the signal-to-noise ratio that future iterative internal lensing reconstructions will achieve as a function of the resolution and sensitivity of CMB temperature and polarization maps. The predicted values of $S/N_\phi$ should be regarded as a lower limit since they were obtained from a minimum variance quadratic estimator and not a MAP reconstruction. The reconstruction noise spectra for the different quadratic estimators were calculated using \verb|quicklens|\footnotemark, and then combined to form the noise spectrum of the minimum variance quadratic estimator following~\cite{qe_formulas}. Large-scale B-modes ($\ell<300$) are excluded from the reconstruction to avoid biasing the delensed B-modes used to iteratively improve the lensing potential estimate (see section~\ref{sec:delensing_methodology}).}
\end{figure}

The estimate of the lensing potential that can be recovered with a minimun variance quadratic estimator~\cite{qe_formulas} is essentially the true angular power spectrum of the lensing potential plus some reconstruction noise:
\begin{equation}
\langle \phi_{LM}^{MV*}\phi_{L'M'}^{MV}\rangle = \delta_{LL'}\delta_{MM'}[C_{L}^{\phi\phi}+N_{L}^{MV}].
\end{equation}
The $N_L^{MV}$ reconstruction noise can be calculated analytically for any cosmological model and experimental configuration (namely the instrumental noise and resolution). The quality of the reconstruction can then be asserted through the signal-to-noise ratio:
\begin{equation}\label{eq:sigmas_de_la_reconstruccion_de_phi}
    S/N_\phi=\left[\sum_{L=2}^{L_{max}}\left(\cfrac{C_L^{\phi\phi}}{C_L^{\phi\phi}+N_L^{MV}}\right)^2\left(L+\frac{1}{2}\right)\right]^{1/2},
\end{equation}
where the maximum available multipole ($L_{max}$) is determined by the map resolution. A forecast on the signal-to-noise that will be achieved from future CMB maps, as a function of their resolution and sensitivity, is shown in figure~\ref{fig:sigmas_of_the_lensing_potential_reconstruction}. These results correspond to an iterative implementation of a minimum variance quadratic estimator. It is noteworthy to remark the fundamental role that the resolution of CMB maps plays in the lensing potential reconstruction, since for a fixed sensitivity, an improvement on resolution can lead to an increase in signal-to-noise of a factor of $\sim 19$ for high $\sigma_n$ (ratio between the 1 arcmin and 30 arcmin curves of figure~\ref{fig:sigmas_of_the_lensing_potential_reconstruction}). As a reference, the Planck Collaboration obtained a $40\sigma$ detection when combining the internal reconstruction coming from their full-mission results with the CIB~\cite{planck2018lensing}.\\

Although introduced here due to its role in the lensing/delensing process, the lensing potential is an extremely interesting observable by itself, since it constitutes an excellent probe for the matter distribution in the universe, going up to much higher redshifts than conventional galaxy surveys. Amongst other applications (see science goals pursued by~\cite{simons_observatory, s4, pico, core_lensing}), a faithful estimate of the lensing potential would help calibrate cluster masses to improve the interpretation of galaxy cluster surveys~\cite{cluster_masses_w/_lensing, hu_cluster_masses, cluster_masses_statistical_approach, cluster_masses_measurements}, provide a measurement of the absolute mass scale of neutrinos~\cite{planck2018lensing, neutrinos_and_cosmology}, and, cross-correlated with other large-scale structure probes, help to achieve a more precise tomographic view of the growth of structure and the cosmic expansion history \cite{giannantonio, bianchini}.\\


\section{Lensing algorithms} 
\label{sec:lensing_algorithms}

\footnotetext{\url{https://github.com/dhanson/quicklens}}

\begin{figure}[tbp]
\centering 
\includegraphics[width=0.6\textwidth]{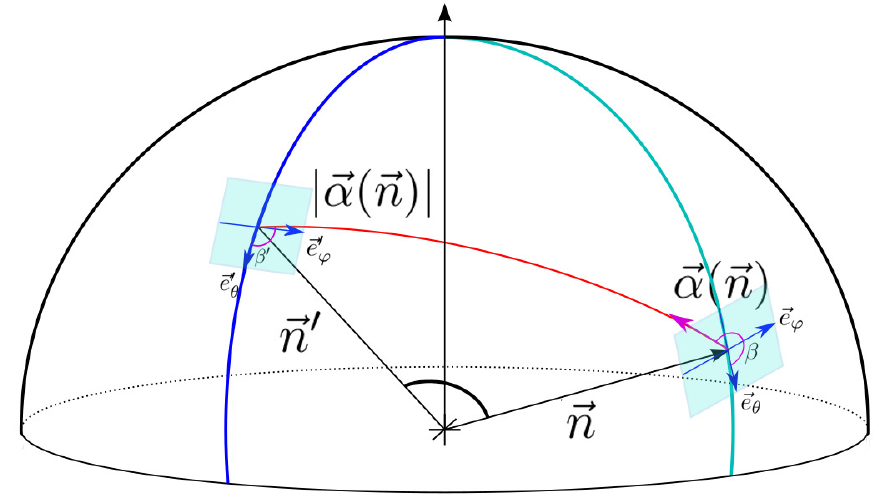}
\hfill
\caption{\label{fig:geometria_lensing_remapping} The original direction $\vec{n}'$ from where a CMB photon was coming can be obtained from the observed direction $\vec{n}$ by moving its end on the surface of the sphere a distance $\alpha=|\vec{\alpha}(\vec{n})|$ along the geodesic in the direction of $\vec{\alpha}(\vec{n})$. In this sketch, deflection angles are blown out of proportion to help visualization.}
\end{figure}

Because of the deflections in their direction of flight that CMB photons suffer while traversing the large-scale structure present in the late universe, the photons we see coming from an $\vec{n}=(\theta,\varphi)$ direction were originally coming from a $\vec{n}'=(\theta ',\varphi +\Delta\varphi)$ direction. Thus, an observed lensed CMB field $\tilde{X}$ (whether it is a temperature or polarization map, i.e., $X=T,Q,U$) is just a remapped version of the original field:
\begin{equation}\label{eq:lensing_remapping}
    \tilde{X}(\vec{n})=X(\vec{n}')=X(\vec{n}+\vec{\alpha}(\vec{n})).
\end{equation}
The original direction can be obtained from the observed direction by moving its end on the surface of the sphere a distance $\alpha=|\vec{\alpha}(\vec{n})|$ along the geodesic in the direction of $\vec{\alpha}(\vec{n})$ (see figure~\ref{fig:geometria_lensing_remapping}). Adopting spherical coordinates, this means that $\vec{n}'$ can be calculated like~\cite{lenspix, lens2hat, das_polynomial_lensing}:\\
\begin{eqnarray}
    \cos \theta '=\cos \alpha \cos \theta -\sin \alpha \sin \theta \cos \beta ; \\
    \sin\Delta\varphi=\cfrac{\sin\beta\sin\alpha}{\sin\theta '},
\end{eqnarray}
where $\beta$ is the angle between the deflection vector $\vec{\alpha}$ and the unitary vector $\vec{e}_\theta$ at position $\vec{n}$.  For polarization, an additional rotation taking into account the different orientation of the basis vectors at the two points must also be applied in order to ensure the parallel transport of vectors along the sphere. The proper way to apply this rotation to the spin-2 polarization field would be
\begin{equation}
    \tilde{P}(\vec{n})=e^{2i\gamma}P(\vec{n}')=e^{2i\gamma}\frac{1}{2}\left(
    \begin{array}{cc}
    Q(\vec{n'}) & -U(\vec{n'})\sin \theta '\\
    -U(\vec{n'})\sin \theta '& -Q(\vec{n}')\sin^2\theta'
    \end{array}
    \right),
\end{equation}
where $\gamma$, the difference between the angles formed between $\vec{e}_\theta$ and the geodesic connecting the two points at $\vec{n}$ and $\vec{n}'$ ($\gamma=\beta-\beta'$ in figure~\ref{fig:geometria_lensing_remapping}), is calculated like
\begin{eqnarray}
    A = \tan \beta ' = \cfrac{\alpha_\varphi}{\alpha\sin\alpha\cot\theta+\alpha_\theta\cos\alpha};\\
    e^{2i\gamma} = \cfrac{2(\alpha_\theta+\alpha_\varphi A)^2}{\alpha^2(1+A^2)} -1 + \cfrac{2i(\alpha_\theta+\alpha_\varphi A)(\alpha_\varphi-\alpha_\theta A)}{\alpha^2(1+A^2)}.
\end{eqnarray}

This $\vec{n}'=\vec{n}+\vec{\alpha}(\vec{n})$ remapping falls under the \emph{Born approximation}, which assumes that lensing deflections can be calculated through potential gradients along the unperturbed path~\cite{lewis&challinor_lensing_review}. Lensing calculations beyond this approximation must include higher order terms of the gravitational potential (as defined in \eqref{eq:total_deflection_angle}, deflections are only computed to first order of $\Phi$), which have the effect of adding a curl component to $\vec{\alpha}(\vec{n})$, and follow the three-dimensional trajectory of photons as they traverse the large-scale structure of the universe to account for individual deflections. When compared to full ray-tracing along the deflected photon path in $N$-body simulations~\cite{ray-tracing_explanation, comparison_ray-tracing_codes}, the Born approximation only starts to fail below arcminute scales ($\ell\geq10^4$), where the weak lensing regime is no longer suitable, and the effects of lens-lens correlation and the non-linearity of the matter distribution become significant~\cite{ray-tracing_explanation, post-born&lens-lens_correlation}. Post-Born terms are also known to have a strong impact in other observables~\cite{ray-tracing_explanation, lensing_bias_cross-correlation_LSS}, such as galaxy-galaxy lensing, higher order statistics of the lensing potential, or CMB lensing cross-correlations. Recent works \cite{bias} predict that although noise levels $\ll 1\mu K\cdot$arcmin would be necessary for detecting the effects of the curl component of $\vec{\alpha}(\vec{n})$, the impact that post-Born corrections have on the lensing potential will already be observable in Stage-IV data. Given the effect they have on the lensing potential, post-Born terms will eventually condition the delensing process. However, determining the impact of post-Born corrections on lensing/delensing is outside of the scope of this work, and we will ignore them here.\\

However, to reproduce the effects of lensing via remapping we must overcome the practical complications that arise when working with discrete maps. When lensing a discrete map via remapping, the value that the lensed map takes at the pixel centered at position $\vec{n}$ corresponds to the value that the unlensed map had at position $\vec{n}+\vec{\alpha}(\vec{n})$, a point that most probably does not correspond with any point in the grid. In principle, we could calculate the exact value that the unlensed field takes at the displaced position from its spherical harmonics like $X(\vec{n}+\vec{\alpha}(\vec{n}))=\sum_{\ell}\sum_m a_{\ell,m}Y_{\ell,m}(\vec{n}+\vec{\alpha}(\vec{n}))$. Since we have to calculate the entire set of $Y_{\ell,m}$ functions at position $\vec{n}+\vec{\alpha}(\vec{n})$, and then sum over them, the computational cost of calculating $X(\vec{n}+\vec{\alpha}(\vec{n}))$ for a single direction is of $\mathcal{O}(\ell_{max}^2)$. Repeating that operation for all the pixels in the lensed map will amount then to $\mathcal{O}(\ell_{max}^2N_{pix})= \mathcal{O}(N_{pix}^2)$\cprotect\footnote{$\ell_{max}\propto N_{pix}^{1/2}$ for the \verb|HEALPix| pixelization.}. Therefore, calculating the exact value that the unlensed field takes at the displaced positions is prohibitive for the map resolutions interesting for lensing ($N_{pix}\sim 10^{7-8}$ pixels), and we must instead approximate $X(\vec{n}+\vec{\alpha}(\vec{n}))$ from the sparse information provided by neighboring pixels. In this way, the problem of remapping becomes a problem of interpolation. The various solutions that have been proposed (and in most cases, implemented in publicly available codes) to solve this problematic are summarized in table~\ref{tab:lensing_algorithms} and will be briefly reviewed in the following subsections.\\

\begin{table}[tbp]
\centering
\begin{tabular}{l p{11cm}}
\hline
\multicolumn{2}{c}{Lensing implementations in the Born approximation}\\
\hline

\multicolumn{2}{l}{Interpolation in an over-sampled grid }\\

\verb|lenS|$^2$\verb|HAT|$^1$~\cite{lens2hat} & Nearest grid point assignment in an over-sampled ECP grid. \\

\verb|LensPix|$^2$~\cite{lenspix, lenspix2} & \verb|Fortran| implementation of a bicubic interpolation in an over-sampled ECP grid.\\

\verb|lenspyx|$^3$ & \verb|Python| implementation of a bicubic interpolation in an over-sampled ECP grid. \\

\verb|FLINTS|$^4$~\cite{flints} & Statistical interpolation that exploits the known properties of isotropic Gaussian fields and the CMB angular power spectra to calculate the value of unlensed fields at any given point from those of the $N$-nearest neighboring pixels.\\

NISHT~\cite{hirata_polynomial_lensing, das_polynomial_lensing} & Lagrange polynomial interpolation in an over-sampled ECP grid. The resampling of CMB fields is accelerated by recasting spherical harmonics as regular Fourier modes. \\

NFFT~\cite{torus} & Unlensed CMB fields are sampled directly at the displaced positions through non-equispaced fast Fourier transforms. \\
\hline

\multicolumn{2}{l}{Taylor expansion}\\
\verb|taylens|$^5$~\cite{taylens} & Approximate unlensed fields at unknown displaced positions $\vec{n}+\vec{\alpha}(\vec{n})$ by its Taylor expansion up to $N$-order around the known pixel center $\vec{n}+\vec{\alpha}_0(\vec{n})$, such that $X(\vec{n}+\vec{\alpha}(\vec{n}))\approx X((\vec{n}+\vec{\alpha}_0(\vec{n}))+\Delta\vec{\alpha}(\vec{n}))$. \\
\hline

\multicolumn{2}{l}{ODE lensing}\\
\verb|LenseFlow|$^6$~\cite{lenseflow} & The lensing operation is translated into an ordinary differential equation by introducing an artificial ``time'' variable to connect lensed and unlensed fields.\\
\hline

\end{tabular}
\cprotect\caption{\label{tab:lensing_algorithms}With the exception of \verb|LenseFlow|, all the algorithms listed are implemented on the sphere. Ray-tracing implementations are not considered here for working beyond the Born approximation (see~\cite{comparison_ray-tracing_codes} for a comparison of ray-tracing codes). Some of the codes are available at the following public repositories:\hfill\break
\footnotesize{$^1$\url{http://www.apc.univ-paris7.fr/APC_CS/Recherche/Adamis/MIDAS09/software/s2hat/vs2hat.html}\\
$^2$\url{https://cosmologist.info/lenspix/}\\
$^3$\url{https://github.com/carronj/lenspyx/tree/master/lenspyx}\\
$^4$\url{http://www2.iap.fr/users/lavaux/software/flints.html}\\
$^5$\url{https://github.com/amaurea/taylens}\\
$^6$\url{https://github.com/marius311/CMBLensing.jl}}}
\end{table}

\subsection{Interpolation in an over-sampled grid}
\label{sec:interpolation_in_an_oversampled_grid}

The most immediate solution to the remapping problem would be to use the $N$-nearest neighboring pixels to interpolate the value of the unlensed map at the displaced position, an idea that has inspired a whole family of lensing implementations. To avoid the severe pixelization errors that would still affect the remapping if a crude nearest neighbor interpolation was used in the \verb|HEALPix|\footnote{\url{http://healpix.sourceforge.net}} grid (the \emph{Hierarchical Equal Area iso-Latitude Pixelation}~\cite{healpix} of the sphere commonly used in CMB science), CMB fields must also be computed at very high angular resolutions.\\

Like previously discussed, the computational cost of working on high resolution pixelizations is very demanding. One of the most extended work-arounds to alleviate this problem is to sample CMB fields in an \emph{equidistant cylindrical projection} of the sphere (ECP, or equirectangular projection~\cite{map_projections}) instead of a \verb|HEALPix| pixelization. In this pixelization, grid points (i.e., pixel centers) are arranged in equidistant iso-latitude rings, with points also equidistant along each ring~\cite{lens2hat}, creating a more interpolation-friendly grid while over-sampling the sphere in comparison to \verb|HEALPix|. Moreover, calculating the value of a field at a given position from its spherical harmonics is three times faster in an ECP pixelization than in the \verb|HEALPix| one. For this reason, many lensing implementations have chosen to internally work in an ECP grid. Starting from the spherical harmonics of a given unlensed CMB field, the unlensed map is then sampled and remapped in this ECP grid, using the interpolation of choice to compute the values at the displaced positions that fall outside of the grid. Finally, the resolution of the resulting lensed map is downgraded to fit into the desired \verb|HEALPix| pixelization. This common structure is shared by most lensing implementations, each of them offering a different take on either the interpolation scheme or the choice of over-sampling grid.\\

In the simplest approach, and if CMB fields are over-sampled up to high enough resolutions, maps can be accurately remapped by just assigning to $\tilde{X}(\vec{n})$ the value that the unlensed map takes at the nearest pixel center to $\vec{n}+\vec{\alpha}(\vec{n})$, without recurring to any interpolation whatsoever. To be computationally affordable in the management of such over-sampled grids, implementations of this algorithm require the use of scalable spherical harmonic transforms and an efficient parallelization of memory, like the one that the \verb|lenS|$^2$\verb|HAT|~\cite{lens2hat} library offers.\\

The over-sampling requirement can be relaxed a little by choosing an appropriate interpolation to complement it. This is precisely what the popular \verb|LensPix|~\cite{lenspix,lenspix2} code does, using, in particular, a modified bicubic interpolation. \cite{flints} suggested a more sophisticated interpolation: the \emph{Fast and Lean Interpolation on the Sphere} (or \verb|FLINTS|). \verb|FLINTS| proposes a fast pixel-based method (without any spherical harmonic algorithm involved) that exploits the known spectral properties of isotropic Gaussian fields to statistically determine the weighting coefficients for the interpolation.\\

The implementation of the aforementioned algorithms can become quite computationally demanding due to the costly summation of spherical harmonics necessary to fill the over-sampled grid. Noting that a band-limited signal in spherical harmonics can be recasted as a band-limited signal in regular Fourier modes on the $(\theta, \varphi)$ plane,~\cite{hirata_polynomial_lensing} proposed a faster resampling of the unlensed fields through two-dimensional Fourier transforms. A Lagrange polynomial interpolation scheme of arbitrary order and precision is then used to compute the values of the field at points outside the chosen ECP grid. This algorithm, known as \emph{non-isolatitude spherical harmonic transform} (NISHT), was succesfully implemented by~\cite{das_polynomial_lensing} to reproduce the lensing of temperature fields in a multi-plane ray-tracing scheme. However, it has not been tested in the lensing of polarization fields. Going a step further,~\cite{torus} suggested to discard the interpolation and directly sample the unlensed fields at the displaced positions using non-equispaced fast fourier transforms (NFFT).\\

\subsection{Taylor expansion}
\label{sec:taylor_expansion}

Instead of interpolating the value that unlensed fields take at displaced positions from that of their neighboring pixels, another option would be to calculate $X(\vec{n}+\vec{\alpha}(\vec{n}))$ through a Taylor expansion of a small $\vec{\alpha}(\vec{n})$ displacement around $X(\vec{n})$. This alternative was first dismissed by \cite{lenspix, taylor_expansion_slow_convergence} for its slow convergence and inefficient scaling with the order of the Taylor expansion, but \cite{taylor_expansion_trick_proposal} later improved its accuracy and efficiency with a simple trick that drastically improved the convergence rate of the expansion.\\

The main drawback hindering the expansion is the fact that although deflection angles are small (typically of a few arcminutes), such displacements are still relatively large compared to the scales involved in the map. The solution to this problem is to divide deflections in two segments: a first $\vec{n}+\vec{\alpha}_0(\vec{n})$ displacement leading to the point in the grid closest to the actual deflected position, and a second smaller $\Delta\vec{\alpha}$ displacement to complete the deflection like $\vec{n}+\vec{\alpha}(\vec{n})= \vec{n}+\vec{\alpha}_0(\vec{n})+\Delta\vec{\alpha}$. Therefore, lensed fields can be calculated through a Taylor expansion up to the $i$-order around the known grid point $\vec{n}+\vec{\alpha}_0(\vec{n})$ like~\cite{taylens}
\begin{equation}
    \tilde{X}(\vec{n})=X((\vec{n}+\vec{\alpha}_0(\vec{n}))+\Delta\vec{\alpha}(\vec{n}))=\sum_{i,j\leq i}\cfrac{\Delta\alpha_\theta^j\Delta\alpha_\varphi^{i-j}}{j!(i-j)!}\partial_\theta^j\partial_\varphi^{i-j}X(\vec{n}+\vec{\alpha}_0(\vec{n})),
\end{equation}
with the later addition of the proper rotation factor necessary for polarization fields. This fragmentation of deflections has the effect of increasing the accuracy of the Taylor expansion as the displacement from the known grid point diminish, and also reducing the order at which the expansion can be truncated (an expansion up to third-order is usually enough to achieve a percent level accuracy in the angular power spectra of lensed fields across the multipole band conceded by the map resolution).\\ 

\subsection{ODE lensing}
\label{sec:ode_lensing}

\cite{lenseflow} proposed a completely different approach to lensing. In the \verb|LenseFlow| algorithm, an auxiliary ``time'' variable $t\in [0,1]$ is introduced to connect the lensed and unlensed fields,
\begin{equation}
    X_t(\vec{n})\equiv X(\vec{n}+t\vec{\alpha}(\vec{n})),
\end{equation}
so that $X_0(\vec{n})=X(\vec{n})$ and $X_1(\vec{n})=\tilde{X}(\vec{n})$. Differentiating over $t$, we can define the homogeneous ordinary differential equation (ODE)
\begin{equation}\label{eq:lenseflow_ode}
    \cfrac{dX_t(\vec{n})}{dt}=
    \frac{\partial X_t(\vec{n})}{\partial n_j} 
    \left[\delta^{ij}+t\cfrac{\partial^2\phi(\vec{n})}{\partial n_i\partial n_j}\right]^{-1}
    \frac{\partial\phi(\vec{n})}{\partial n_i},
\end{equation}
where $\delta^{ij}$ is the Kronecker delta running over the $\vec{n}=(x,y)$ variables in the plane (\verb|LenseFlow| has yet to be implemented in the sphere). In this way, the lensing operation has been transformed from a remapping of points across the sphere into the solving of an ODE problem for each pixel, starting from initial conditions $X(\vec{n})$.\\

This innovative approach to lensing offers several advantages, the most relevant of them (at least for this work) being that within this formulation the lensing operation is exactly reversed by simply running the ODE backward in time (i.e., $t=1\rightarrow 0$ against the $t=0\rightarrow 1$ direction of lensing). Therefore, the only limiting factor in the accuracy of the lensing/delensing operation is the ODE time-step discretization error. The invertibility of \verb|LenseFlow| also extends to individual pixel-to-pixel lensing (by replacing the derivatives in \eqref{eq:lenseflow_ode} with their discrete Fourier analog), achieving a numerically stable and accurate lensing even for relatively large pixels.\\


\section{Comparison of delensing methodologies}
\label{sec:delensing_methodologies_general}

One of the core goals of this project was to compare different delensing methodologies with the objective of finding which one would be the optimal to apply in the analysis of the data to come in the next decade. Even more importantly, we also wanted to determine if current tools would be able to fully delens said data or, on the contrary, the development of new delensing methodologies was needed. For this purpose, we will first review the procedure prescribed by each methodology in section \ref{sec:delensing_strategies}, and then, in section~\ref{sec:delensing_performance_comparison}, compare the delensing efficiency that can be reached when implementing them with a code from each of the three distinct families of lensing algorithms presented in section~\ref{sec:lensing_algorithms}. We will compare their performance first in an ideal noiseless case, and then in the typical conditions expected for future CMB experiments.\\

\subsection{Delensing strategies}
\label{sec:delensing_strategies}

Conceptually, reverting the effect of lensing should be as simple as remapping the observed photons back to their original positions. Therefore, the underlying unlensed field could be recovered from an observed lensed field through a new deflection $\vec{\beta}(\vec{n})$ in the same way as the lensed field was obtained by remapping points with $\vec{\alpha}(\vec{n})$:
\begin{equation}\label{eq:inverse/lensing_displacements}
    \tilde{X}(\vec{n})=X(\vec{n}+\vec{\alpha}(\vec{n})) \leftrightarrow X(\vec{n})=\tilde{X}(\vec{n}+\vec{\beta}(\vec{n})).
\end{equation}
Remembering the definition of $\vec{\alpha}(\vec{n})$, we can think of this inverse displacement as the gradient of an \emph{inverse} projected mass distribution so that $\vec{\beta}(\vec{n})=\vec{\nabla}\phi^{inv}(\vec{n})$. The new deflection can be determined by imposing the condition that points are remapped onto themselves after being deflected back and forth,
\begin{equation}\label{eq:points_remapped_onto_themselves}
    \vec{n}+\vec{\beta}(\vec{n})+\vec{\alpha}( \vec{n}+\vec{\beta}(\vec{n}))=\vec{n},
\end{equation}
so that
\begin{equation}\label{eq:inverse_deflection}
 \vec{\beta}(\vec{n})=-\vec{\alpha}(\vec{n}+\vec{\beta}(\vec{n}))= -\vec{\nabla}\phi(\vec{n}+\vec{\beta}(\vec{n})).
\end{equation}
According to this definition, the inverse displacement is then a warped version of a curl-free vector field. This warping acts as a rotation, introducing an additional non-zero divergence-free term to the inverse displacement. To explicitly acknowledge the existence of this term, we could write the Helmholtz decomposition of the inverse displacement like~\cite{hirata&seljak, likelihood_reparameterization}
\begin{equation}\label{eq:inverse_potential_all_terms}
    \vec{\beta}(\vec{n})=\vec{\nabla}\phi^{inv}(\vec{n})+ \star \vec{\nabla}\psi^{inv}(\vec{n}),
\end{equation}
where $\star$ represent the $90^\circ$ rotation of the $\psi^{inv}$ stream function potential used to model a field rotation. The term  $\star\vec{\nabla}\psi^{inv}(\vec{n})$ is present even if the forward displacement is a pure gradient (as assumed in~\eqref{eq:total_deflection_angle}). However, the contribution of such curl component can be safely ignored until estimates of the lensing potential reach sub-percent accuracy levels according to~\cite{likelihood_reparameterization}.\\

\begin{figure}[tbp]
\centering 
\includegraphics[width=0.9\textwidth]{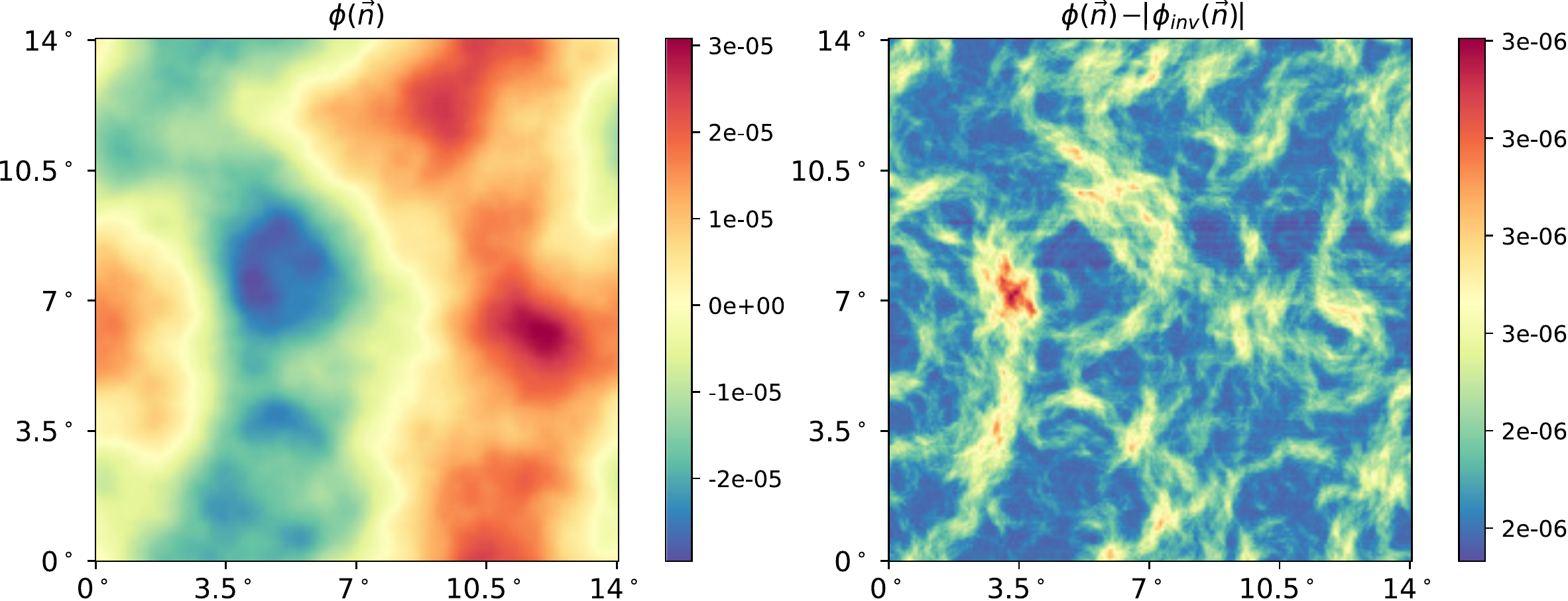}
\hfill
\cprotect\caption{\label{fig:comparacion_phi_vs_phi_inv} Difference between the lensing and inverse lensing potentials (right panel) for a given random realization of $\phi(\vec{n})$ in the plane (shown in the left panel for reference). At first order, $\phi^{inv}(\vec{n})$ is approximately $-\phi(\vec{n})$, so differences between the lensing and inverse lensing potential are calculated like $\phi(\vec{n})-|\phi^{inv}(\vec{n})|$. The inverse lensing potential is constructed from the inverse displacement defined in~\eqref{eq:inverse_deflection} to satisfy $\vec{\beta}(\vec{n})=\vec{\nabla}\phi^{inv}(\vec{n})$. }
\end{figure}

Adopting a Newton-Raphson scheme, the inverse deflection field can then be iteratively calculated through~\cite{planck_first_internal_delensing_temperature&polarization, lensit}:
\begin{equation}\label{eq:iterative_inverse_displacement}
    \vec{\beta}_{i+1}(\vec{n}) = \vec{\beta}_i(\vec{n}) - M^{-1}(\vec{n}+\vec{\beta}_i(\vec{n})) \left[ \vec{\beta}_i( \vec{n}) + \vec{\alpha}(\vec{n}+\vec{\beta}_i(\vec{n})) \right],
\end{equation}
where $M$ is the magnification matrix, defined from the sphere's metric $g_{ab}$, and the covariant derivatives of the deflection like
\begin{equation}\label{eq:magnification_matrix}
    M_{ab}(\vec{n})=g_{ab}+\nabla_a\alpha_b(\vec{n}).
\end{equation}
Starting from a $\vec{\beta}_0(\vec{n})=0$ deflection, and working in sub-arcmin grids on the plane, equation \eqref{eq:iterative_inverse_displacement} rapidly leads to a stable solution for $\phi^{inv}(\vec{n})$\footnote{Since lensing codes actually take the potential as their input, the convergence of equation \eqref{eq:iterative_inverse_displacement} must be evaluated in terms of $\phi^{inv}_i(\vec{n})$ instead of $\vec{\beta}_i(\vec{n})$ to also include the small numerical errors that the integration from $\vec{\beta}_i(\vec{n})$ to $\phi^{inv}_i(\vec{n})$ introduces and thus fully asses the progress of the iterative calculation.} after only four iterations for the typical $\Lambda$CDM deflection field. Here we considered $\phi^{inv}_i(\vec{n})$ to have fully converged when the RMS of the $\phi_i^{inv}(\vec{n}) + \phi(\vec{n})$ difference stabilizes and the delensing that we can achieve with $\phi_i^{inv}(\vec{n})$ no longer improves with further iterations. \\

Remapping with the inverse deflection should be the exact way of reverting lensing (up to the ignored $\star\vec{\nabla}\psi^{inv}(\vec{n})$ term in \eqref{eq:inverse_potential_all_terms}). On the other hand, it would require the solving of equation~\eqref{eq:iterative_inverse_displacement} on the full sphere at arcmin resolutions (meaning $\sim 10^{7-8}$ pixels in a \verb|HEALPix| grid), a task computationally consuming given how fields have to be interpolated at unknown positions $\vec{n}+\vec{\beta}_i(\vec{n})$ in every iteration. Alternatively, seeing how differences between the lensing and inverse lensing potential are small (see figure~\ref{fig:comparacion_phi_vs_phi_inv}), one could adopt the \emph{antilensing approximation}, where the inverse deflection is simply taken to be $\vec{\beta}(\vec{n})\approx -\vec{\alpha}(\vec{n}$) and thus $\phi^{inv}(\vec{n})\approx-\phi(\vec{n})$.\\

However, given how the lensing kernel operates on polarization fields, a remapping of points (either with $\vec{\beta}(\vec{n})$ or directly $-\vec{\alpha}(\vec{n})$) is not the optimal approach to delensing B-modes. At its very core, lensing is an intrinsically different operation for E- and B-mode polarization~\cite{lewis&challinor_lensing_review, hu_lensing_harmonic_approach, lens2hat}: the lensing of E-modes is a self-contained operation, where for any given multipole $\ell_i$ the main contribution for the $C_\ell^{\phi\phi}\otimes C_\ell^{EE}$ convolution comes from a narrow band centered around $\ell_i$; meanwhile, lensed B-modes are entirely created by the leakage of E-modes, with all scales in the $C_\ell^{\phi\phi}\otimes C_\ell^{EE}$ convolution presenting an equally relevant contribution to the total power of the lensed angular power spectrum at $\ell_i$. In this way, if we envision the lensing operation as a matrix product between angular power spectra, the matrix responsible for the lensing of B-modes will strongly couple all the scales of the potential and E-mode spectra, while the matrix responsible for the lensing of E-modes will be fairly diagonal.\\

\begin{figure}[tbp]
\centering 
\includegraphics[width=0.6\textwidth]{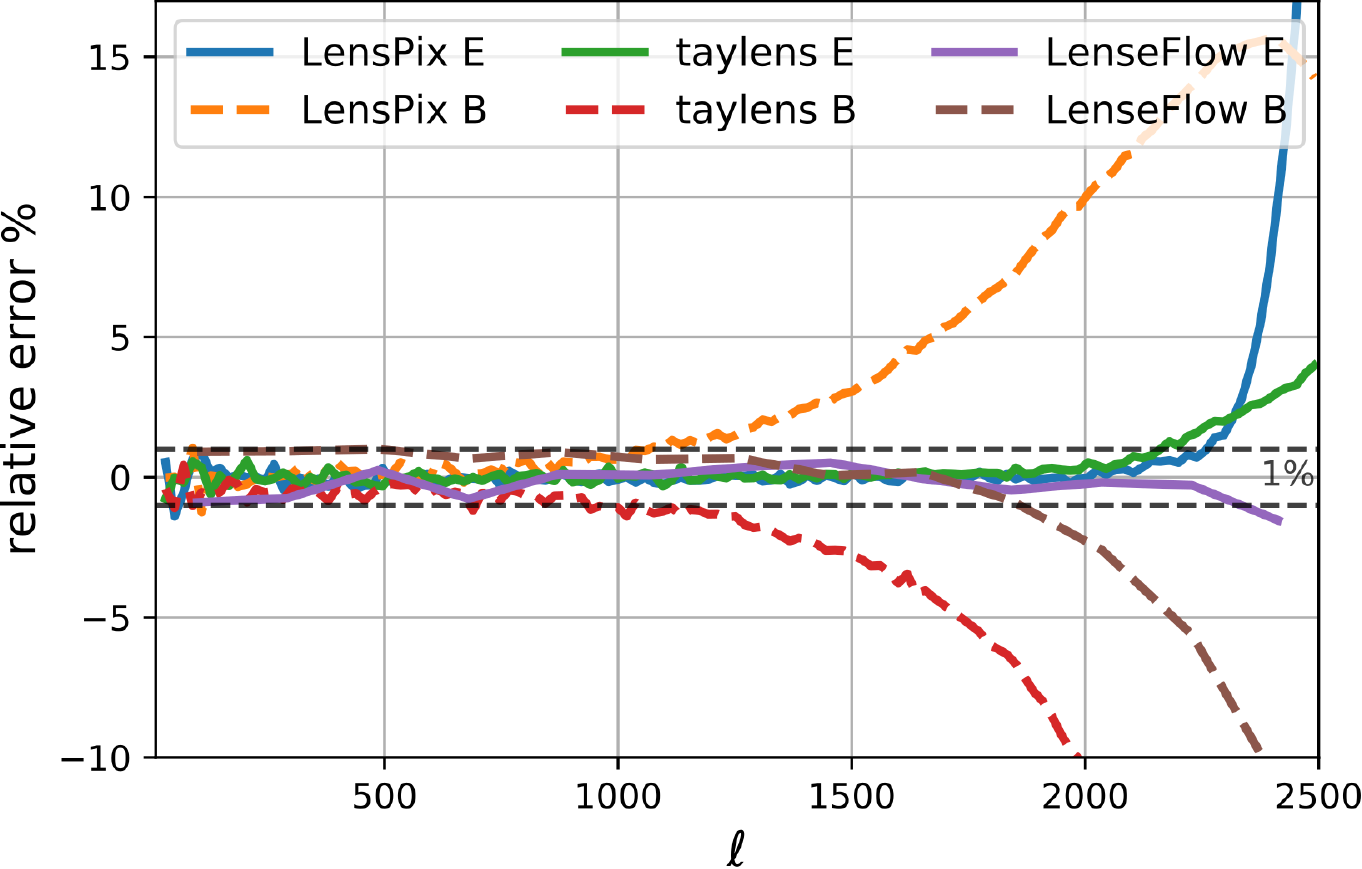}
\hfill
\cprotect\caption{\label{fig:lens_accuracy_E_vs_B} Relative error $(1-C_{\ell,lens}/C_{\ell,input})\times 100$ in the reproduction of lensed E- and B-modes obtained when lensing noiseless $3.4$ arcmin resolution maps with a representative code from each of the three families of algorithms presented in table~\ref{tab:lensing_algorithms}. Having fixed the resolution of the maps to ensure a fair comparison, we chose the optimal values for the tuning parameters of each code that were recommended in their respective documentations. Spectra are conveniently binned and averaged over 25 simulations.}
\end{figure}

As a consequence, when working with real-space maps, the accuracy of the lensing operation will always be limited by the multipole range that the map resolution can accommodate, with this band limitation having a more severe impact on the lensing of B-modes. To illustrate this point, figure~\ref{fig:lens_accuracy_E_vs_B} shows the relative error with respect to the input\footnote{ We use \texttt{CAMB}~\cite{camb} to produce the input unlensed and lensed CMB angular power spectra. Hence, in this comparison of map-based lensing implementations, we are implicitly assuming that \texttt{CAMB}'s lensed $C_\ell$ correspond to the true lensed angular power spectra. A discussion on the validity of that assumption is outside of the scope of this work.} angular power spectra in the reproduction on lensed E- and B-mode fields obtained when lensing noiseless $3.4$ arcmin resolution maps with a code from each of the three families presented in table~\ref{tab:lensing_algorithms}. For the two codes implemented in the sphere (\verb|LensPix| and \verb|taylens|), the lensing of E-modes is well behaved until almost reaching the map resolution limit, while the error in the reproduction of lensed B-modes starts to explode much earlier. These restrictions that band-limitation imposes on the accuracy of lensed B-modes at high $\ell$ can be partially ameliorated by using a more sophisticated interpolation scheme (like, for example, the \verb|FLINTS|~\cite{flints} interpolation) to compensate the lack of information at the smaller scales. On the other hand, \verb|LenseFlow|'s implementation is limited to the plane, which denies the access to the larger scales since we must work with small regions of the sky (in particular, here we are working with $29^\circ\times 29^\circ$ patches). Therefore, although it reports a better accuracy than \verb|LensPix| or \verb|taylens| in the reproduction of lensing at the smaller scales ($\ell>1000$), \verb|LenseFlow| systematically underestimates lensed B-modes at the larger scales.\\

Acknowledging this band limitation, a better delensing should be achieved by concentrating lensing operations on the E-mode channel. With this in mind, an alternative delensing procedure will be to produce a template of the expected B-modes by lensing the estimate of the unlensed E-mode (obtained whether with $\vec{\beta}(\vec{n})$ or $-\vec{\alpha}(\vec{n}$)), and then subtracting it from the observed lensed B-mode map. This methodology, to which we will refer as template delensing, has already been extensively used in many forecasting works (e.g., \cite{delensing_w/_cib, hirata&seljak_eb_qe, delensing_w/_external_datasets}) and even applied to real data~\cite{planck2018lensing, delensing_sptpol_w/_herschel, template_delensing_polarbear}. Although counter-intuitive at first, we will see how the introduction of the additional lensing operation in the construction of the template does indeed improve the final degree of delensing because it allows us to avoid the strong coupling of scales between E-modes and the lensing potential that would be severely affected by the band limitation in a direct delensing of B-modes.\\

Lastly, the \verb|LenseFlow| implementation offers an exact way of reverting lensing on a pixel-by-pixel scale, since once lensing is understood as an ODE problem, delensing is trivially done by just running the ODE in reverse. We will use this exact delensing to validate the results obtained with the other methodologies. Conceptually, this operation should be equivalent to remapping with the inverse potential, but we will see how this is not the case.\\

\subsection{Performance comparison}
\label{sec:delensing_performance_comparison}

\begin{figure}[tbp]
\centering 
\includegraphics[width=0.6\textwidth]{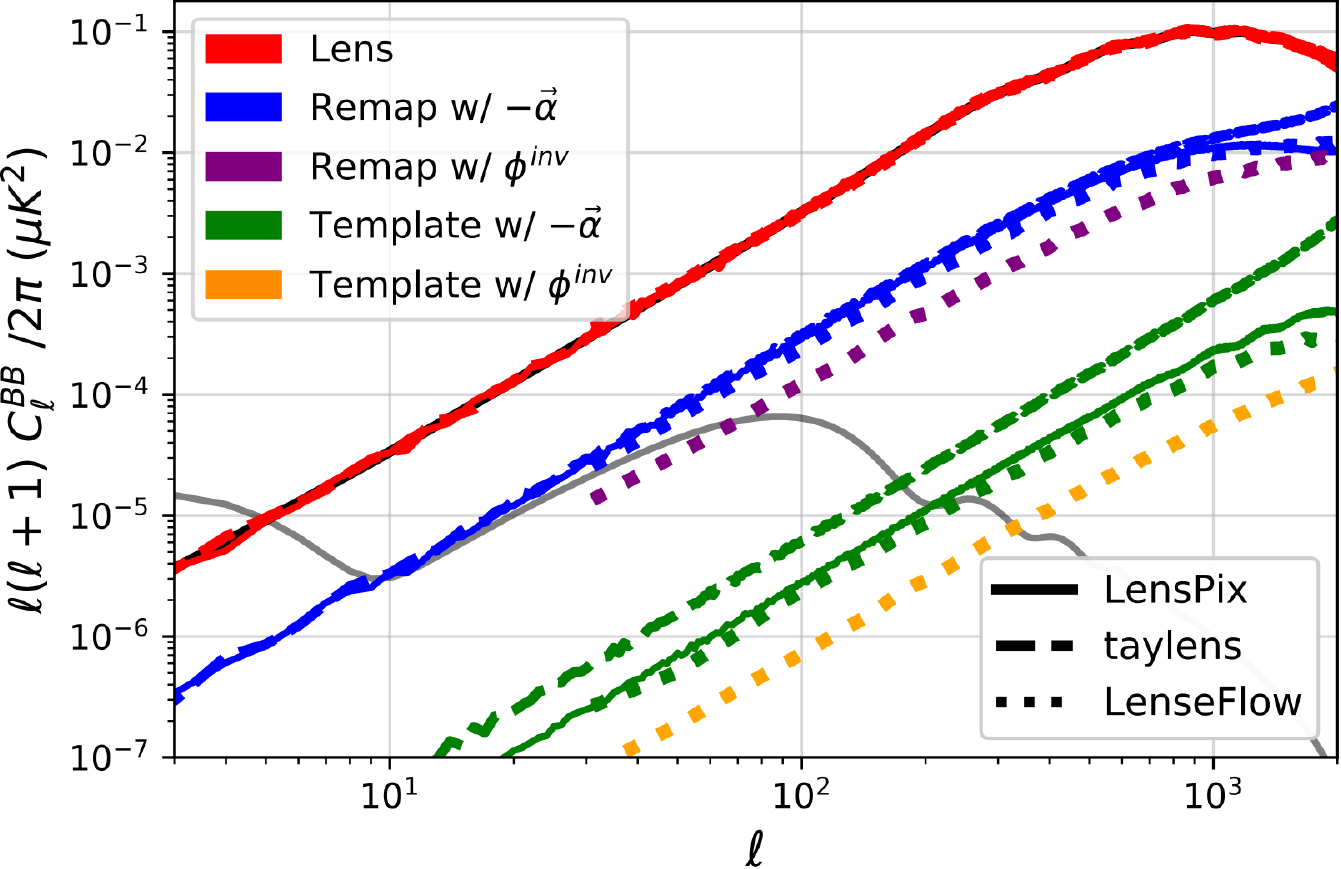}
\hfill
\cprotect\caption{\label{fig:comparacion_delensings} Comparison of the performance of the different delensing methodologies when implemented with a representative code from each of the three families of algorithms presented in table~\ref{tab:lensing_algorithms}. The type of line (solid, dashed or dotted) indicates what code was used in each case. Like in figure~\ref{fig:lens_accuracy_E_vs_B}, here we are working with noiseless $3.4$ arcmin resolution maps. Starting from the lensed map each code produced (red lines), we delens it within the antilensing approximation, either through a plain remapping (blue lines), or through the construction of a template of the expected lensed B-modes (green lines). In addition, since working in the plane vastly reduces computational costs, we also test remapping (purple line) and template delensing (orange line) with the inverse lensing potential with \verb|LenseFlow|. Because its implementation is limited to the plane, we do not have access to the lowest multipoles when using \verb|LenseFlow| (we are working with $29^\circ\times29^\circ$ sky patches). Spectra are conveniently binned and averaged over 25 simulations.}
\end{figure}

We proceed now to compare the performance of the delensing methodologies presented in the previous section. For this purpose we lensed a set of CMB simulations with their corresponding lensing potentials, and then delensed them applying the different methodologies, repeating the operation with a code from each of the three families of lensing algorithms. The results of this exercise are displayed in figure \ref{fig:comparacion_delensings}. For computational convenience, delensing with the inverse displacement was only implemented in the plane. The delensing efficiency will be quantitatively evaluated in terms of the \emph{delensing fraction}, defined like
\begin{equation}\label{eq:delensing_fraction_definition}
    \mathcal{D}=\langle C_{\ell, delens}^{BB}/C_{\ell, lens}^{BB}\rangle_{\ell\leq200}.
\end{equation}
Hence, the better the delensing is, the lower the delensing fraction we will get (ideally we want $\mathcal{D}=0$).\\

As figure~\ref{fig:comparacion_delensings} shows, there is a clear hierarchy between methodologies when working in the ideal scenario of knowing both the exact potential and CMB (i.e., working with noiseless maps). Like anticipated, template delensing proves to be much more efficient than plain remapping, whether it be in the antilensing approximation ($\mathcal{D}_{remap}/\mathcal{D}_{template}\approx 110$) or using the inverse potential ($\mathcal{D}_{remap}/ \mathcal{D}_{template} \approx 170$). Fixing instead the chosen methodology, we can also quantify that using the inverse potential is a factor of $\approx 2$ better than using the antilensing approximation for plain remapping, and a factor of $\approx 3$ better for template delensing.\\

In addition, the accuracy to reproduce the lensing effect of each algorithm (see figure~\ref{fig:lens_accuracy_E_vs_B}) further conditions the delensing efficiency. The \verb|LenseFlow| implementation reports the best results, proving to be around a $20\%$ more efficient than \verb|LensPix|, both for remapping and template delensing. In comparison, lensing via Taylor expansion turns out to be the worst approach, with \verb|taylens| reporting a delensing fraction $\approx 1.25$ times worse than that of \verb|LenseFlow| for remapping, and $\approx 2.71$ times worse when it comes to template delensing. Looking at figure~\ref{fig:comparacion_delensings}, the template delensed spectrum obtained with \verb|taylens| seems to behave more like noise than a lensed signal, suggesting that, within this implementation, enough numerical noise might be accumulated in the extra lensing step needed for template construction to significantly affect the overall delensing efficiency. Attending to the systematically better performance \verb|LenseFlow| has demonstrated, the implementation of an ODE lensing algorithm for the full sphere would be very interesting.\\

\begin{figure}[tbp]
\centering 
\includegraphics[width=1.0\textwidth]{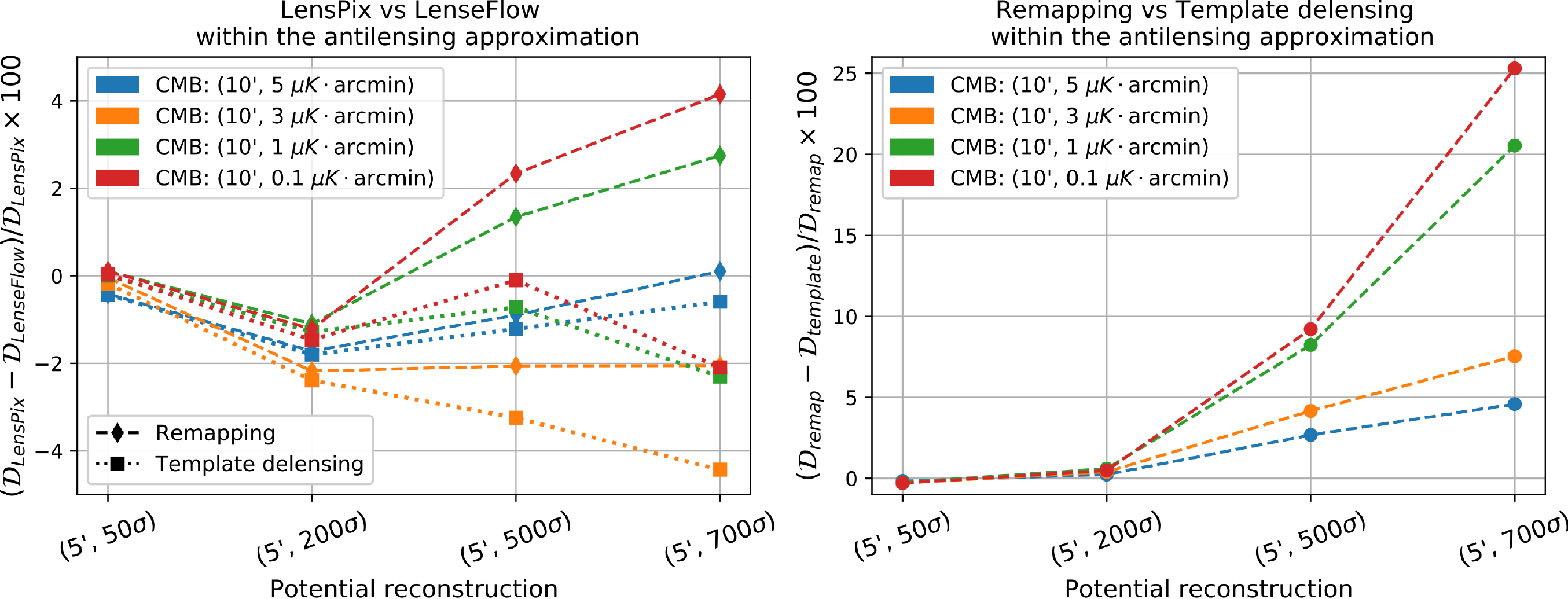}
\hfill
\cprotect\caption{\label{fig:comparacion_delensing_fractions} Comparison between the delensing fractions that could be recovered when applying different delensing methodologies and implementations to the typical noisy maps expected for next-generation experiments. The x-axis runs over increasingly better lensing potential reconstructions, while the color code indicates the kind of noisy CMB map assumed in each case. \emph{Left)} Discrepancy between delensing fractions recovered with \verb|LensPix| and \verb|LenseFlow| for both a plain remapping (dashed diamonds) and a template delensing (dotted squares) within the antilensing approximation. For results to be comparable, the same pixel resolution and multipole binning were used in both implementations.  \emph{Right)} Comparison of the delensing fractions recovered with \verb|LenseFlow| when choosing plain remapping and/or template delensing (both within the antilensing approximation) as the delensing methodology. Template delensing will only start to report a significantly better delensing efficiency than plain remapping for good enough potential reconstructions and low-noise CMB maps.}
\end{figure}

Leaving behind such ideal scenario, the performance of the different methodologies and implementations starts to converge when working with the kind of noisy maps that may be expected from next-generation experiments. To prove this point, we downgraded the resolution and added isotropic Gaussian noise to CMB and lensing potential maps after lensing them, and then proceeded to delens such noisy CMB maps with those degraded lensing potentials. The comparison between the ideal scenario and this more realistic case allows us to conclude several points. First of all, like the left panel of figure~\ref{fig:comparacion_delensing_fractions} shows, the delensing efficiencies obtained with \verb|LenseFlow| and \verb|LensPix| become very similar (differences below $4.5\%$) now that we are working with noisy maps. In fact, most of the points in that graph are negative, indicating that \verb|LensPix| tends to have a sightly better performance than \verb|LenseFlow|. The only exceptions are those cases with exceptionally good potential reconstructions and very low instrumental noises. Based on the little difference between them, and the lesser computational cost of working with small patches in the plane, we would interchangeably use one or the other implementation for our forecast. The little discrepancy between the two algorithms in realistic situations also alleviates the urgency of taking \verb|LenseFlow| to the sphere, since equivalent results can be obtained with already available implementations.\\

Let us remark that the benefit that the use of $\vec{\beta}(\vec{n})$ instead of $-\vec{\alpha}(\vec{n})$ reported for template delensing also gets reduced in the presence of noise, converging both methodologies down to sub-percent discrepancies for all the combinations of CMB and potential reconstructions we tested. Based on this good agreement, we can safely adopt template delensing within the antilensing approximation as the default procedure, avoiding the computationally expensive step of calculating the inverse displacement without any loss of delensing efficiency. The same cannot be said for remapping, where the use of the inverse displacement will still report a substantial benefit with respect to antilensing when applied to low-noise CMB maps and good enough potential reconstructions.\\

Furthermore, a sub-precent agreement is also found when comparing template delensing within the antilensing approximation with the exact lensing inversion of running the \verb|LenseFlow| ODE in reverse. This validation against the exact solution, in addition with the previous point, allows us to affirm that template delesing within the antilensing approximation will be the optimal delensing methodology to employ in the analysis of the data to come in the next decade since it reduces the computational time with no significant loss of accuracy. Most importantly, the good agreement between template delensing and the exact solution tells us that currently available implementations and known methods are already capable of fully exploiting the delensing possibilities for next-generation experiments.\\ 

Advocating for computational convenience, the right panel of figure~\ref{fig:comparacion_delensing_fractions} demonstrates how, for poor potential reconstructions, one could simply apply a remapping within the antilensing approximation, avoiding the extra lensing step of constructing the template, without much loss of delensing efficiency.\\

\section{Delensing in next-generation skies}
\label{sec:predicciones_generales}

In this section we aim to determine how well it would be possible to revert lensing on the B-mode polarization maps that the next generation of CMB experiments will provide, drawing especial attention to the dependence that delensing efficiency has with the properties of the input CMB maps and lensing potential reconstructions. For this purpose, we will simulate CMB maps with the typical instrumental specifications of next-generation experiments, and lensing potential reconstructions of the typical signal-to-noise that could be expected, and apply to them the delensing techniques studied in section \ref{sec:delensing_methodologies_general}. We will explain our simulation pipeline and delensing methodology in subsection \ref{sec:delensing_methodology}, and present our results in subsection \ref{sec:predictions}.\\

\subsection{Simulations and delensing methodology}
\label{sec:delensing_methodology}

First of all, we need to build a dataset of CMB and lensing potential simulations representative of the data to come in the following years. To emulate future experimental situations, we will add isotropic Gaussian noise and downgrade the resolution of both lensed CMB and lensing potential maps to match the typical instrumental specifications for next-generation CMB experiments. Here we will consider polarization fields and lensing maps to be unrelated entities (as if the potential reconstruction was provided by an external agent instead of internally produced from each particular CMB realization), thus avoiding the additional complications of contemplating the existing correlations between the two fields~\cite{point_out_delensing_bias, characterize_delensing_bias, delensing_bias_temperature, planck_first_internal_delensing_temperature&polarization, namikawa,bias}.\\

However, this assumption can not be made when the estimate of the potential is internally produced from the CMB itself, like it happens with the iterative reconstructions used in sections \ref{sec:lensing_potential} and \ref{sec:detectability}. In those scenarios, we adopt the simple method proposed by \cite{bias} to avoid biasing the delensing due to the correlations that arise between B-modes and the potential reconstruction when B-modes are used to estimate $\phi$: remove the multipoles of interest for delensing (namely $\ell <300$) from the B-mode map entering the potential reconstruction. By applying this methodology we obtain an unbiased delensed B-mode, at the cost of a small degradation on the signal-to-noise of potential reconstructions. The removal of large-scale B-modes has a negligible effect on the signal-to-noise of potential reconstructions coming from CMB maps of high resolution (only a $\sim 1\%$ loss in $S/N_\phi$ for maps of a $5$ arcmin resolution and $\lesssim 2\mu K\cdot$arcmin sensitivities) since there is a great deal of small-scale information beyond $\ell=300$. It is only for CMB maps of low resolution, like for example those of LiteBIRD-like experiments ($\sim 30$ arcmin resolution), that the $\ell<300$ cut will result in a significant reduction of the reconstrution's signal-to-noise: around a $\sim 40\%$ loss in signal-to-noise for a $2\mu K\cdot$ arcmin sensitivity, and a $\sim 20\%$ loss for a $0.5\mu K\cdot$ arcmin sensitivity. Based on the typical resolutions and sensitivities expected for the next generation of CMB experiments, this loss in signal-to-noise would only lead to a $\lesssim10\%$ degradation of the delensing fraction.\\

For the next generation of CMB experiments, the typical angular resolution of cosmological channels ranges between $20$ to $5$ arcmin, and instrumental sensitivities fall around the $\mu K\cdot$arcmin scale~\cite{simons_observatory, s4, pico, litebird}. Hence we will consider polarization maps of a $20$, $10$ and $5$ arcmin resolution with instrumental noises of $5$, $3$, and $1$ $\mu K\cdot$arcmin. As a lower limit, we will also test a $0.1$ $\mu K\cdot$arcmin noise. Looking at the forecast on figure~\ref{fig:sigmas_of_the_lensing_potential_reconstruction}, lensing potential reconstructions of signal-to-noise ratios of $200$ and $500$ could be produced from next-generation data. For completeness, we will also consider a $S/N_\phi=50$ reconstruction, a case close to the currently available Planck $40\sigma$ detection~\cite{planck2018lensing}, and the upper limit case of a $S/N_\phi=700$ reconstruction. The $S/N_\phi$ associated to any given reconstruction is translated into a white noise $N_L$ angular power spectrum through equation~\eqref{eq:sigmas_de_la_reconstruccion_de_phi}, and finally projected into the real-space variance of an isotropic Gaussian noise realization. We will fix a $5$ arcmin resolution for lensing potential maps.\\ 

\begin{figure}[tbp]
\centering 
\includegraphics[width=0.8\textwidth]{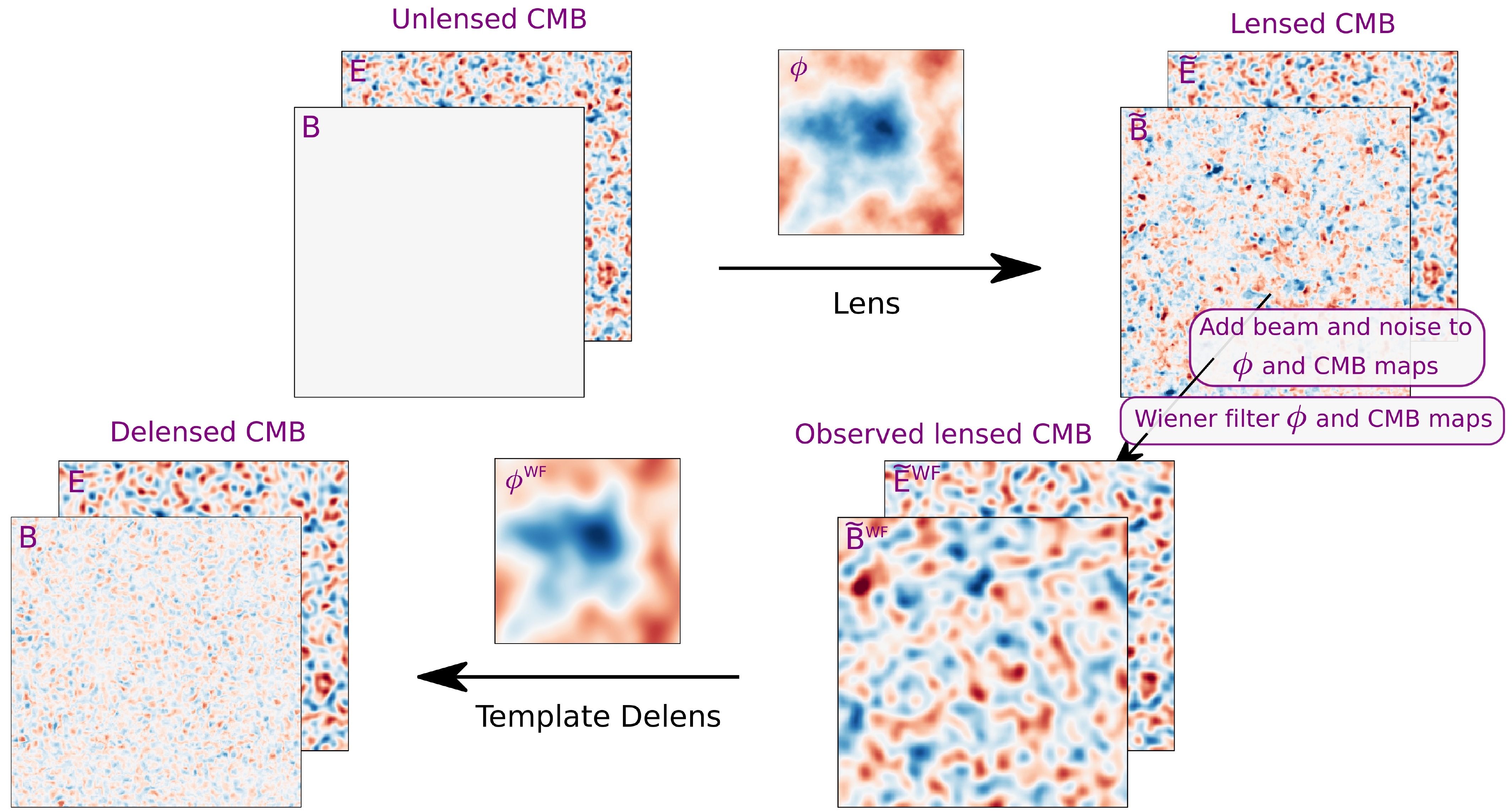}
\hfill
\caption{\label{fig:simulation_pipeline} Flow diagram summarizing the simulation pipeline for the delensing process. First, a random realization of CMB and lensing potential maps is computed. After lensing the CMB, the resolution of both lensing potential and lensed CMB maps is downgraded, and a realization of isotropic white Gaussian noise is added. Both maps are then Wiener filtered to counteract the weight of the noise dominated scales. Finally, the simulated observed lensed CMB maps are template delensed using the fictitious lensing potential reconstruction.}
\end{figure}

With the dataset of observed CMB maps and lensing reconstructions ready, the first leg of our simulation pipeline is completed (see figure~\ref{fig:simulation_pipeline}). Now we will define our delensing methodology. In section~\ref{sec:delensing_performance_comparison}, we proved that template delensing within the antilensing approximation on the full-sky with \verb|LensPix| and on the plane with \verb|LenseFlow| presents less than a $4.5\%$ discrepancy in the recovered delensing fraction for the typical maps that next-generation CMB experiments will provide. Based on this good agreement, and for the sake of computational convenience, we will then adopt template delensing with $-\vec{\alpha}(\vec{n})$ on the plane as our delensing methodology. Although we will not be able to probe the largest scales when constrained to the plane, we can safetly extrapolate the recovered $\mathcal{D}$ to the lowest multipoles.\cprotect\footnote{Another way to put it is that we will be using \verb|LenseFlow| in the plane to approximate the result \verb|LensPix| will return in the full-sky, where we know that the $C_{\ell,delens}^{BB} = \mathcal{D} C_{\ell,lens}^{BB}$ holds for all $\ell<100$.} Delensing fractions will be calculated as the quotient between template delensed and lensed angular power spectra like shown in \eqref{eq:delensing_fraction_definition}.\\

We found the Wiener filtering of lensed maps to be a fundamental step prior to any delensing attempt. Intuitively, the importance of this filtering process can be understood as a necessity to make the observed map as faithful to the underlying lensed field as possible, since the subsequent delensing will completely rely on the assumption that lensed fields are just a remapped version of the unlensed ones. As the addition of instrumental beam and noise distort the original lensed fields, the \eqref{eq:inverse/lensing_displacements} equivalence starts to fail, and observed fields receive the delensing remapping as a new lensing. As a consequence, without the filtering process, observed maps would often present a higher lensed B-mode signal after being delensed. However, the very own nature of the lensed B-mode signal (equivalent to a $5$ $\mu K\cdot$arcmin white noise up to $\ell\sim 300$) imposes a restriction on the range of instrumental noises that the Wiener filter can counteract without starting to act as a distortion itself. Therefore, as will be evidenced by our results, a $5$ $\mu K\cdot$arcmin noise is the natural limit at which delensing processes of all kinds would start to fail (e.g., \cite{caldeira} also encountered the $5$ $\mu K\cdot$arcmin limit when applying deep neural networks to the delensing problem).\\

For this reason we will filter, both, the potential, and the E- and B-mode maps before delensing. The filtering window function applied to polarization CMB fields ($X=E,B$) will be:
\begin{equation}\label{eq:wf_cmb}
    \omega_\ell^X=\frac{1}{b_\ell}\frac{C_\ell^{XX}}{C_\ell^{XX}+N_\ell^{XX}b_\ell^{-2}},
\end{equation}
where $b_\ell$ represents the instrumental beam, and $N_\ell^{XX}$ the instrumental white noise added to the simulations. A window function like this one will already reduce the amplitude of B-modes without performing any delensing, especially when dealing with high levels of noise. To distinguish between both effects, in the next section we will consider, both, the equivalent $\mathcal{D}$ achieved by just applying the Wiener filter, and the actual $\mathcal{D}$ obtained after the full delensing process.\\

In turn, the window function applied to potential maps is constructed from the white $N_L$ noise we assumed for these external lensing potential reconstructions:
\begin{equation}
    \omega_L^\phi=\frac{1}{b_L}\frac{C_L^{\phi\phi}}{C_L^{\phi\phi}+N_Lb_L^{-2}}.
\end{equation}
If instead of with these fictitious $\phi$ estimates we were to work with internal reconstructions like the ones used in sections \ref{sec:lensing_potential} and \ref{sec:detectability}, the window function would be defined from the $N_L^{MV}$ noise of the minimum variance quadratic estimator like:
\begin{equation}
    \omega_L^{MV}=\frac{C_L^{\phi\phi}}{C_L^{\phi\phi}+N_L^{MV}}.
\end{equation}
In this case, the effect of the instrumental beam is already accounted for (in a non-trivial way, see e.g. \cite{qe_formulas}) in the reconstruction noise.\\

\subsection{Dependence of delensing efficiency on data quality}
\label{sec:predictions}

\begin{figure}[tbp]
\centering 
\includegraphics[width=1.0\textwidth]{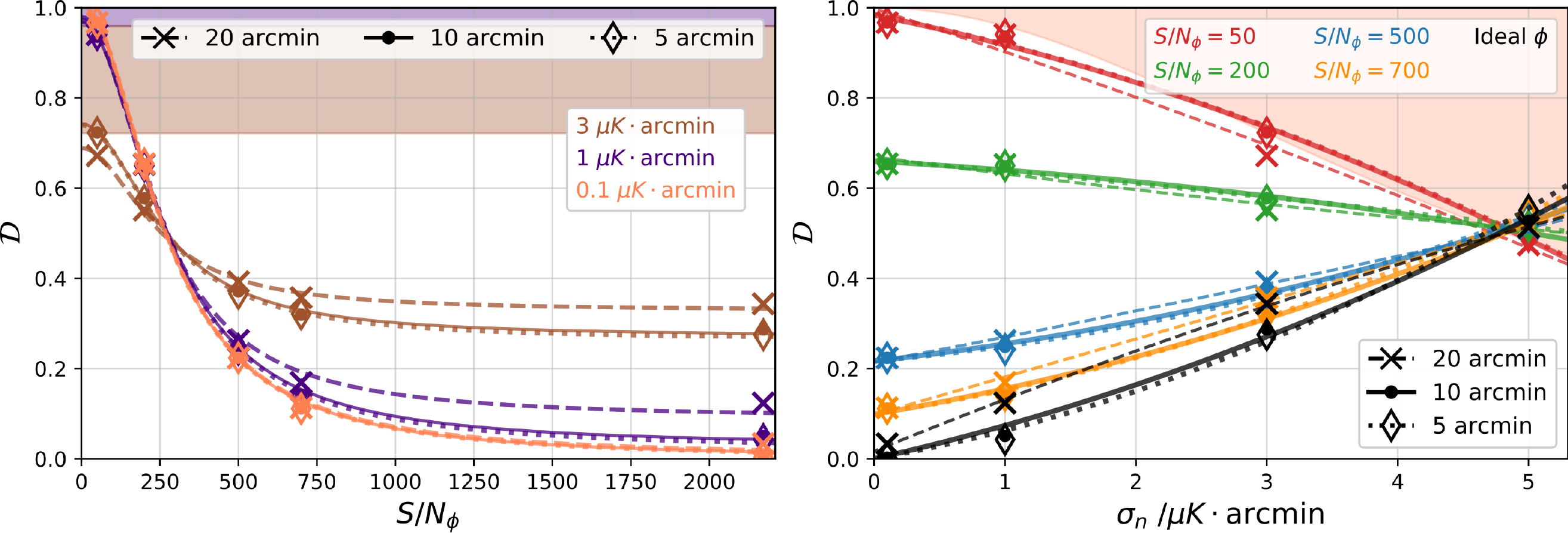}
\hfill
\caption{\label{fig:template_delens_D_vs_sigma} Delensing fraction as a function of the quality of the lensing potential reconstruction (left), and as a function of the instrumental sensitivity (right) for the different instrumental specifications explored. Diamonds and dotted lines correspond to a CMB map resolution of $5$ arcmin, while dots and solid lines are associated to a $10$ arcmin resolution, and crosses and dashed lines show the results for a $20$ arcmin resolution. Markers tend to overlap because the resolution of the CMB map only has a significant effect on the delensing fraction for low $\sigma_n$ and high $S/N_\phi$. The ideal situation of delensing with the exact lensing potential is represented by black curves in the right panel, and by points at $S/N_\phi=2172$ (cosmic variance limit of equation~\eqref{eq:sigmas_de_la_reconstruccion_de_phi}) in the left panel. The lower boundary of the colored areas marks the fraction of B-modes that is reduced by applying the Wiener filter defined in \eqref{eq:wf_cmb}. Therefore, for all points enclosed in those colored regions, delensing will not reduce B-modes further than what the use of the Wiener filter already would. For a sensitivity of $0.1\mu K\cdot$arcmin, the window function of the Wiener filter is very close to unity, and hence its correspondent colored region is unnoticeable in the panel on the left. }
\end{figure}

To understand how delensing efficiency depends on the quality of the data it is performed on, we followed the simulation pipeline sketched in figure~\ref{fig:simulation_pipeline} to obtain delensed maps with the aforementioned experimental specifications, averaging angular power spectra over $25$ simulations before calculating delensing fractions as in~\eqref{eq:delensing_fraction_definition}. The recovered delensing fractions are shown in figure~\ref{fig:template_delens_D_vs_sigma}, both as a function of the quality of the lensing potential reconstruction (characterized by the $S/N_\phi$), and the instrumental sensitivity.\\

Like previously mentioned, Wiener filtering is a necessary step before delensing, but filtering with a window function like \eqref{eq:wf_cmb} will already reduce the amplitude of B-modes without having performed any delensing whatsoever. For this reason, we must compare the obtained delensing fractions with the reduction of B-modes associated to Wiener filtering in order to distinguish which fraction of B-modes were removed due to the filtering, and which fraction was actually a consequence of the delensing. To allow this comparison, the lower boundary of the colored areas in figure~\ref{fig:template_delens_D_vs_sigma} marks the reduction of B-modes achieved through Wiener filtering for the different instrumental sensitivities considered. In this way, for every delensing fraction plotted in those graphs, the fraction of $\mathcal{D}$ inside the colored region shows the reduction of B-modes that is a product of the filtering, and the remaining fraction outside of the colored region shows the reduction of B-modes that is purely a consequence of delensing. Therefore, delensing will not reduce B-modes further than what the use of the Wiener filter already did for all the points enclosed in the colored regions. This includes all CMB maps with instrumental sensitivities $\geq5\mu K\cdot$arcmin.\\

In addition to the $5\mu K\cdot$arcmin limiting case, figure \ref{fig:template_delens_D_vs_sigma} shows how Wiener filtering would be the main contribution to B-mode reduction for high levels of noise and poor potential reconstructions. For the poorest of reconstructions explored, the $S/N_\phi=50$ case, the effect of delensing would be negligible even for the best of instrumental sensitivities. For intermediate $S/N_\phi=200$ reconstructions, Wiener filtering would be the main contribution to $\mathcal{D}$ for noises $\sigma_n\gtrsim 3\mu K\cdot$arcmin, while delensing will start to dominate the reduction of B-modes below that threshold. Finally, delensing will be the primary cause for the reduction of B-modes when good enough reconstructions of the lensing potential, $S/N_\phi\gtrsim 500$, are available. In this way, and although it is not evident at a first glance,  delensing efficiency always improves towards lower instrumental sensitivities in the right panel of figure \ref{fig:template_delens_D_vs_sigma} once the contribution of the Wiener filter is subtracted.\\

Another two noteworthy remarks can be drawn from a quick analysis of the graphs in figure~\ref{fig:template_delens_D_vs_sigma}. First of all, as black curves (on the right panel) and points at $S/N_\phi=2172$ (on the left panel) showcase, even in the ideal situation of delensing with the exact lensing potential, the minimum delensing fraction we could reach is already limited by the instrumental noise and resolution of the observed CMB. Hence the selection of a target delensing fraction should be considered as another driver for functional requirements when designing a new CMB experiment. The second feature to highlight is the small impact that instrumental resolution has on the delensing fraction (overlapping of lines and markers of the same color) when working within the $\sigma_n<3\mu K\cdot$arcmin and $S/N_\phi\leq 500$ range. Although not surprising in the upgrade from $20$ to $10$ arcmin beams since both of them are way broader than the average lensing displacement ($\langle|\vec{\alpha}|\rangle\sim 2$ arcmin), the increase of resolution from $10$ to $5$ arcmin could have led to an appreciable effect as we started to approach the scale at which we could be sensitive to individual displacements. This independence with instrumental resolution contrasts with the extreme sensitivity that internal lensing potential reconstruction presents towards it (see figure~\ref{fig:sigmas_of_the_lensing_potential_reconstruction}). Therefore, optimal experiment design for delensing and for potential reconstruction do not necessarily coincide and should be considered separately.\\

\begin{table}[tbp]
\centering
\begin{tabular}{|c|ccc|}
\hline
\diagbox{$S/N_\phi$}{ FWHM} &  $20$ arcmin &  $10$ arcmin &  $5$ arcmin\\
\hline
50 & $( -2.6, -9.3, 10.0 )$ & $( -8.3, -5.8, 9.8 )$ & $( -8.1, -5.9, 9.9 )$ \\
200 & $( 1.4, -3.9, 6.7 )$ & $( -3.1, -1.7, 6.6 )$ & $( -1.1, -2.4, 6.6 )$ \\
500 & $( 0.9, 5.5, 2.1 )$ & $( 5.9, 3.2, 2.2 )$ & $( 7.8, 2.5, 2.2 )$  \\
700 & $( 0.0, 8.4, 1.0 )$ & $( 7.0, 5.0, 1.0 )$ & $( 9.0, 4.4, 1.0 )$ \\
Ideal & $( -3.8, 11.9, 0.2 )$ & $( 8.1, 6.6, 0.0 )$ & $( 12.0, 5.1, 0.0 )$ \\
\hline
\end{tabular}
\caption{\label{tab:D(sigma_n)_fit} $(a\times 10^3,b\times 10^2, c\times 10^1)$ parameters for every combination of potential reconstruction and FWHM of the CMB beam obtained when fitting the points shown in the right panel of figure \ref{fig:template_delens_D_vs_sigma} with a $\mathcal{D}(\sigma_n)=a\sigma_n^2+b\sigma_n +c$ law. Thus, the $c$ parameter is dimensionless, and $a$ and $b$ have units of, respectively, ($\mu K\cdot$arcmin)$^{-2}$ and ($\mu K\cdot$arcmin)$^{-1}$.}
\end{table}

\begin{table}[tbp]
\centering
\begin{tabular}{|c|ccc|}
\hline
\diagbox{ $\sigma_n$}{ FWHM} & $20$ arcmin & $10$ arcmin & $5$ arcmin\\
\hline
 3 $\mu K\cdot$arcmin & $( 2.1 , 6.0 , 3.3 )$ & $( 3.3 , 7.0 , 2.7)$ & $( 3.4 , 7.1 , 2.6 )$\\
 1 $\mu K\cdot$arcmin & $( 5.7 , 6.4 , 0.9 )$ & $( 7.1 , 7.6 , 0.3 )$ & $( 7.2 , 7.6 , 0.2 )$ \\
 0.1 $\mu K\cdot$arcmin & $( 7.1 , 7.1 , 0.0 )$ & $( 7.2 , 7.1 , 0.0  )$ & $( 7.2 , 7.2 , 0.0 )$ \\
\hline
\end{tabular}
\caption{\label{tab:D(S/N_phi)_fit} $(a\times 10^{-4},b\times 10^{-4},c\times 10^{1})$ parameters for every combination of CMB beam and instrumental noise obtained when fitting the points shown in the left panel of figure \ref{fig:template_delens_D_vs_sigma} with a $\mathcal{D}(S/N_\phi)=\frac{a}{(S/N_\phi)^2+b}+c$ law. Thus, all three $a$, $b$ and $c$ parameters are dimensionless.}
\end{table}

Quantitatively, the dependence observed in the right panel of figure~\ref{fig:template_delens_D_vs_sigma} can be fitted by a phenomenological $\mathcal{D}(\sigma_n)=a\sigma_n^2+b\sigma_n +c$ curve, with the quality of the potential reconstruction fixing how flat or steep the curve is, and whether delensing fractions will increase or decrease towards lower values of $\sigma_n$. Thus, once the quality of the lensing potential is set (i.e., the values of $a$, $b$, and $c$ are determined), the best delensing efficiency that can be possibly achieved by improving the sensitivity of the CMB experiment is also fixed. The triads of $(a,b,c)$ values we obtained in our fit are shown in table \ref{tab:D(sigma_n)_fit}. We can also use this simple law to predict that future delensing efforts will only prove to be successful, in the sense that a $\mathcal{D}$ below the equivalent delensing fraction associated to Wiener filtering could be reached, for sensitivities below a $\sigma_n \approx 4.5$ $\mu K\cdot$arcmin threshold. Correspondingly, the dependence seen on the left panel of figure~\ref{fig:template_delens_D_vs_sigma} can be approximately reproduced using a $\mathcal{D}(S/N_\phi)= \frac{a}{(S/N_\phi)^2+b}+c$ law with the values for the $(a,b,c)$ parameters shown in table \ref{tab:D(S/N_phi)_fit}.\\

Even in the best of the scenarios considered, delensing fractions do not reach the $\mathcal{D}=7\times 10^{-4}$ floor we saw template delensing presents in ideal conditions (see figure~\ref{fig:comparacion_delensings}). Therefore, the delensing of next-generation data will be limited by the properties of the available maps rather than by the limitations of current delensing methodologies.\\

\section{Implications for PGWB detection}
\label{sec:detectability}

After having studied how instrumental specifications and potential reconstruction quality condition delensing efficiency, we can now predict what delensing fractions could future experiments achieve, and how much delensing would help them to improve their sensitivity to detect the PGWB. First of all, we will set the scene for delensing in the near future in subsection~\ref{sec:cada_experimento_individualmente }, indicating the instrumental specifications assumed for each experiment, and determining the delensing potential of each of them individually. As previously discussed, the optimal experimental configuration for delensing is not necessarily the optimal configuration for lensing potential reconstruction. Therefore, in subsection~\ref{sec:deslensar_litebird_con_ground} we will consider the benefits that delensing the LiteBIRD sky with internal lensing potential reconstructions coming from ground-based experiments could report. Finally, the joint analysis of datasets from the different experiments will be discussed in subsection~\ref{sec:analisis_conjunto}.\\

\subsection{Delensing forecast for future experiments}
\label{sec:cada_experimento_individualmente }

A plethora of new experiments have already been proposed to exploit the information ingrained in the polarization of the CMB during the next decades, including both ground-, like the Simons Observatory (SO)~\cite{simons_observatory} and the CMB Stage-IV (S4)~\cite{s4}, and space-based experiments, like the LiteBIRD~\cite{litebird, litebird_update} and PICO~\cite{pico} satellites. As a summary, table~\ref{tab:experiment_parameters} collects some of their typical instrumental specifications, namely instrumental sensitivity, beam resolution, and sky coverage. Just from these properties alone we could already determine which degree of delensing would be possible given the instrumental constraints by applying once again the procedure explained in section~\ref{sec:delensing_methodology} (see figure~\ref{fig:simulation_pipeline}), or what kind of internal potential reconstruction could be produced by iteratively applying a minimum variance quadratic estimator.\\

However, and although ignored in this work until now, the contribution of Galactic foreground emission also needs to be taken into account to properly evaluate the detectability of the PGWB. Since a proper study of how foregrounds affect the delensing efficiency is beyond the scope of this work, we will not include them in our simulations at the map level. Such a study is left for a future work. Nevertheless, we will approximate their effect at the angular power spectrum level by adding a foreground residual component to the observed B-mode spectra used for both, the internal lensing potential reconstruction, and the evaluation of the sensitivity to PGWB detection. Without delving into the specifics of component separation methods, we will model foreground residuals as a certain $R_F$ fraction of the foreground B-mode spectrum expected at the $1\%$ cleanest fraction of the sky at 100 GHz~\cite{errard} (approximately a $F_\ell=1.5\times 10^{-2}\ell^{-2.29}$ law, see figure~\ref{fig:lensing_dampening_of_acoustic_oscillations}). In this way we mimic a signal that mantains the structure of a foreground signal, thus acting like a true \emph{residual} component rather than a systematic or statistical noise introduced during component separation. \cite{errard} provides a forecast for the foreground residuals of LiteBIRD and the S4 Ultra-Deep (UD) survey, which we then scale by the square of the $\sigma_n$ ratio between experiments to obtain the rest of $R_F$s. To account for the different configurations, we scale ground-based experiments using the S4 Ultra-Deep prediction as a reference, and space-based experiments using the LiteBIRD prediction. Knowing that this is a naive method for calculating $R_F$, we will let the foreground residuals vary between a $[R_F/3, R_F\times3]$ of the tabulated value in the evaluation of the signal-to-noise to account for the uncertainties in our approximate estimation of $R_F$.\\

\begin{table}[tbp]
\centering
\begin{tabular}{|l|ccccc|}
\hline
\multirow{ 2}{*}{Experiment} & $\mathrm{FWHM}$ & $\sigma_n$ &  \multirow{ 2}{*}{$f_{sky}$} & \multirow{ 2}{*}{$\Delta\ell$} & \multirow{ 2}{*}{$R_F$}   \\
& /arcmin &  /$\mu K\cdot$arcmin & &   & \\
\hline

LiteBIRD & 30.0 & 2.0 & 80 & $2<\ell<200$ & 0.01 \\

PICO & 6.2 & 0.87 & 80 & $2<\ell<1000$ & 0.001  \\

SO SAT (@145 GHz) & 17.0 & 2.1 & 10 & $10<\ell<1000$ & 0.05  \\
SO LAT (@145 GHz) & 1.4 & 6.3 & 40 & $10<\ell<1000$ & 0.43 \\

S4 DW (@155 GHz) & 1.4 & 2.8 & 64 & $10<\ell<1000$ & 0.08 \\
S4 UD (@155 GHz) & 1.5 & 0.96 & 3 & $10<\ell<1000$ & 0.01   \\
\hline
\end{tabular}
\caption{\label{tab:experiment_parameters} Instrumental beam, sensitivity and sky coverage assumed for each experiment. For ground-based experiments, we chose the specifications of a representative cosmological channel: 145 GHz for the Simons Observatory~\cite{simons_observatory}, and 155 GHz for the Stage-IV~\cite{s4}. In turn, the $\sigma_n$ adopted for space-based experiments corresponds to the total combined polarization sensitivity over the full mission. Following \cite{litebird, litebird_update}, we take $30$ arcmin to be the typical angular resolution of LiteBIRD at 150 GHz. The chosen resolution for PICO~\cite{pico} is that of its best sensitivity channel (155 GHz). The resolution and sky coverage of each experiment also constrain the $\Delta\ell$ multipole band available for the angular power spectrum analysis. Multipoles beyond $\ell>1000$ are deprecated since they report no significant contribution to the overall signal-to-noise. Foreground residuals remaining after component separation are modeled as $R_F$ times the foreground B-mode spectrum expected for the $1\%$ cleanest fraction of the sky at 100 GHz~\cite{errard}.}
\end{table}

Once all components of the observed B-mode angular power spectrum are identified (the $L_\ell$ lensed B-modes, $F_\ell$ Galactic foregrounds, $N_\ell\omega_\ell^{-2}$ noise deconvolved by the instrumental beam, and the $B_\ell$ primordial B-mode for an $r=1$), we can model the observed signal as a sum of all of them,  and fit for it:
\begin{equation}\label{eq:chi_cuadrado}
    \chi^2(r, A_L, A_F)=\sum_{\ell_1}^{\ell_2}\cfrac{\left(C_\ell^{obs}-rB_\ell-A_LL_\ell-A_FF_\ell-N_\ell\omega_\ell^{-2}\right)^2}{\sigma_\ell^2},
\end{equation}
where the error associated to each multipole is the cosmic variance corresponding to the observed B-mode spectrum (assuming a fiducial $r_{fid}$, a delensing fraction $\mathcal{D}$, and foreground residuals $R_F$)
\begin{equation}\label{eq:cosmic_variance}
    \sigma_\ell^2=\cfrac{(C_\ell^{obs})^2}{f_{sky}(\ell+0.5)}=\cfrac{\left(r_{fid}B_\ell+\mathcal{D}L_\ell+R_FF_\ell+N_\ell\omega_\ell^{-2}\right)^2}{f_{sky}(\ell+0.5)}.
\end{equation}
Looking at these equations, one can easily understand how lensed B-modes limit the detection of a PGWB. The problem is not that we confuse the lensing signal ($L_\ell$ is well-known, given a cosmological model determined by the other CMB angular power spectra, and can be fitted for), but that it introduces an additional contribution to the cosmic variance that acts like an inescapable $5$ $\mu K\cdot$arcmin white noise. Therefore, to achieve a PGWB detection, we have to reduce the lensing signal in \eqref{eq:cosmic_variance} (i.e., delens). From the $\chi^2$ defined in \eqref{eq:chi_cuadrado} we can construct the corresponding Fisher matrix, and estimate from it the $\sigma_r(r)$ uncertainty in the determination of the tensor-to-scalar ratio after marginalizing over $A_L$ and $A_F$. The signal-to-noise of the detection would then be $S/N(r)=r/\sigma_r(r)$.\\

Having explained our forecasting methodology, we can now start to present our predictions. The first thing to notice is that, for the SO LAT survey, none of the studied delensing methodologies would be able to reduce B-modes further than what the use of the Wiener filter already will, since they require noises below the $\sigma_n\lesssim4$ $\mu K\cdot$arcmin threshold to work efficiently (see section~\ref{sec:predictions}). Accordingly, this also means that the internal lensing potential reconstruction cannot be iteratively improved by successively delensing with the recovered potential estimate. Nevertheless, and although we will not be considering it here since that reduction of B-modes is not a direct consequence of delensing, a simple filtering like the one in \eqref{eq:wf_cmb} would be enough to substantially reduce the amplitude of the observed B-modes when dealing with such high levels of noise. Column $\mathcal{D}_{min}$ in table~\ref{tab:delensing_forecast} shows the minimun delensing fraction (i.e., the best delensing possible) that the instrumental constraints of the observed lensed CMB allow.  For the experiments with sensitivities around $2\mu K\cdot$arcmin (LiteBIRD, SO SAT and S4 DW), the best they can aspire to is to reduce the lensing signal down to $\sim 1/5-1/3$. Nevertheless, such delensing efficiencies would still report a significant gain in signal-to-noise, as the $\mathcal{G}^r=S/N(r)_{w/ \hspace{1mm} delens}/S/N(r)_{w/o \hspace{1mm} delens}$ parameter in table~\ref{tab:delensing_forecast} shows. Sensitivities below the $1$ $\mu K\cdot$arcmin level, like the ones sought by PICO or the S4 UD survey, would be necessary to reduce the amplitude of lensed B-modes down to $<5\%$.\\

\begin{table}[tbp]
\centering
\begin{tabular}{|l|cccc|}
\hline
\multirow{ 2}{*}{Experiment} & \multirow{ 2}{*}{$S/N_\phi$} & \multirow{ 2}{*}{$\mathcal{D}_{auto}$} & \multirow{ 2}{*}{$\mathcal{D}_{min}$} & \multirow{ 2}{*}{$\mathcal{G}^{0.001}$}\\
&  &  &  &   \\
\hline

LiteBIRD & 66 & 0.73 & 0.33 & 2.1 \\

PICO & 622 & 0.20 & 0.03 & 9.7 \\

SO SAT (@145 GHz) & 178 & 0.59 & 0.22 & 2.4\\
SO LAT (@145 GHz) & 421 & - & - & -\\

S4 DW (@155 GHz) & 750 & 0.38 & 0.24 & 2.0 \\
S4 UD (@155 GHz) & 1473 & 0.12 & 0.04 & 7.6 \\
\hline
\end{tabular}
\caption{\label{tab:delensing_forecast} Forecast of the delensing potential of next-generation experiments. The quality of the lensing potential reconstruction that could be iteratively produced with a minimum variance quadratic estimator is evaluated through the signal-to-noise ratio $S/N_\phi$ as if all experiments were full-sky (to account for their actual coverage, just multiply the shown values by $\sqrt{f_{sky}}$). For the SO LAT survey, none of the studied delensing methodologies would be able to reduce B-modes further than what the use of the Wiener filter already will, since they require noises below the $\sigma_n\lesssim4$ $\mu K\cdot$arcmin threshold to work efficiently. Hence, the potential reconstruction cannot be iteratively improved either. $\mathcal{D}_{min}$ shows the minimum delensing fraction (i.e., the best delensing possible) allowed by the instrumental beam and sensitivity, and $\mathcal{D}_{auto}$ the one that would be obtained when delensing with a potential of the same $S/N_\phi$ as its own internal reconstruction. To highlight the contribution that delensing can make to PGWB detection, $\mathcal{G}^{0.001}$ shows the fractional gain in signal-to-noise that a $\mathcal{D}_{min}$ delensing would report in an $r=0.001$ scenario.}
\end{table}

Column $\mathcal{D}_{auto}$ shows the delensing fraction that experiments would reach if they were to be delensed by its own internal lensing potential reconstruction. We recall that the $S/N_\phi$ value obtained through a minimum variance quadratic estimator must be regarded as a lower limit to the optimum MAP reconstruction, and that experiments could also further improve their lensing potential estimates by combining their internal reconstruction with other large-scale structure tracers, such as galaxy surveys or the CIB (e.g., as done in~\cite{delensing_sptpol_w/_herschel, planck2018lensing}). For CMB experiments with bands in the $\sim200-800$ GHz range, this could be done without even recurring to external datasets, since they overlap with the frequency range where the CIB can be observed. Thus, there is still room for the improvement of the $\mathcal{D}_{auto}$ values shown in table~\ref{tab:delensing_forecast}. Although still far from reaching the $\mathcal{D}_{min}$ limit imposed by instrumental constraints, PICO and the S4 UD would be able to reduce lensed B-modes down to, respectively, $\sim 1/5$ and $\sim1/10$ of its initial amplitude. In contrast, due to its broad beam, which leads to a poor potential reconstruction, LiteBIRD will no be capable of fully exploiting its delensing capabilities on its own. For this reason, we will entertain the idea of using potential reconstructions coming from ground-based experiments to delens the LiteBIRD sky in the following subsection.\\

\begin{figure}[tbp]
\centering 
\includegraphics[width=1.0\textwidth]{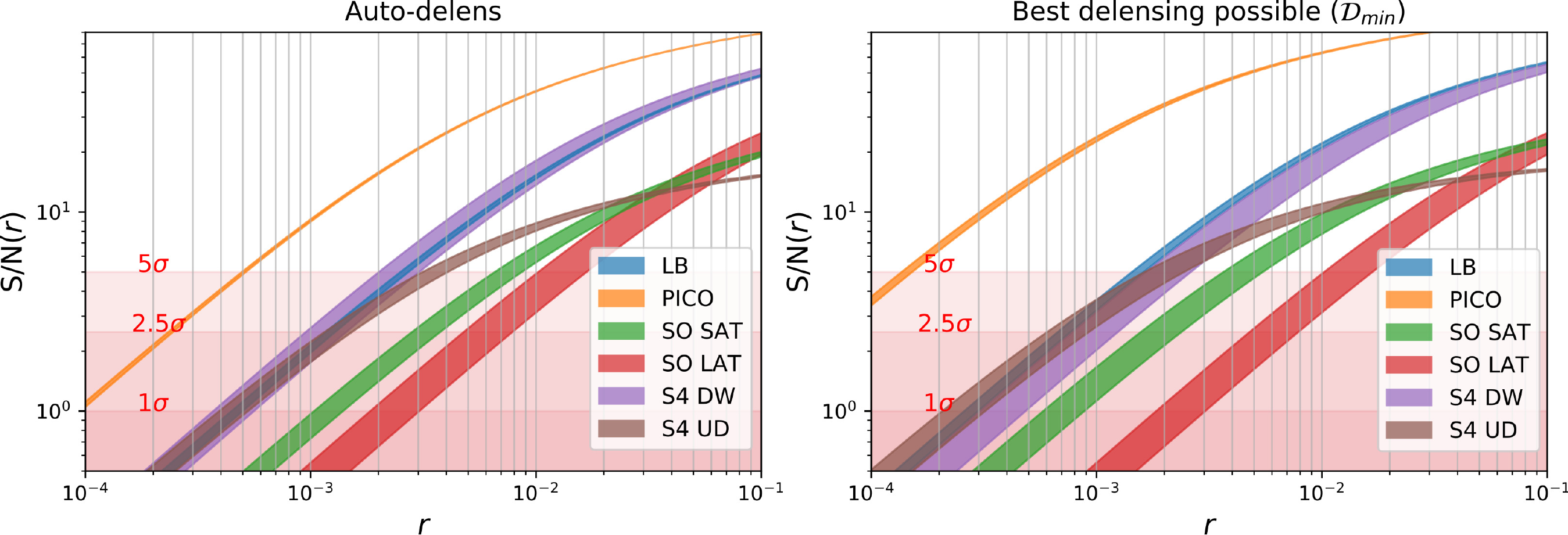}
\hfill
\caption{\label{fig:auto_max_delens_all_exps} Signal-to-noise of the PGWB detection that next-generation experiments would achieve if they could reach the maximum delensing efficiency allowed by their instrumental beam and sensitivity (right), and when delensed with a potential of the same $S/N_\phi$ as its own internal reconstruction (left). Contours span the range of signal-to-noise compatible with an $[R_F/3, R_F\times 3]$ uncertainty in the estimation of the foreground residuals. The results shown for SO LAT are not delensed. }
\end{figure}

The left panel of figure~\ref{fig:auto_max_delens_all_exps} shows the repercussions that this auto-delensing has in terms of signal-to-noise for the different experiments. Thanks to its higher resolutions, and the better delensing that this allows, the S4 can compensate its smaller sky coverage and the loss in signal-to-noise due to not having access to the reionization peak of the primordial B-mode spectrum (i.e., the signal coming from $\ell<10$), resulting in a signal-to-noise comparable to that of a full sky experiment like LiteBIRD. The PICO satellite joins both a full sky coverage with high resolution and low noise, thus becoming the ultimate probe for PGWB detection. In comparison, the intermediate step that SO constitutes between current and future ground-based experiments (a stage-III compared to the desired stage-IV) would only be enough to grant a $2\sigma$ detection of an $r=2.4\times 10^{-3}$.  \\

Like the contours in figure \ref{fig:auto_max_delens_all_exps} reveal, our signal-to-noise predictions for ground-based experiments are more sensitive to the chosen level of foreground residuals than the predictions for space-based experiments. This dependency reflects the fact that, because of their instrumental noise levels and the degrees of delensing they can achieve, foreground residuals are the dominant component in signal-to-noise determination for ground-based experiments. Note that, at least at the level at which we are considering them, the impact that foregrounds have on the potential reconstruction is much less important than the one they have in the actual determination of $r$. Therefore, when we start combining datasets from space- and ground-based experiments in following subsections, the likelihood will be dominated by the term coming from the space-based experiment, and, again, the predicted signal-to-noise of combinations of space- and ground-based experiments will be more robust against the value of $R_F$ than the signal-to-noise of combinations that only involve ground-based experiments.\\

Finally, to stress the valuable contribution that delensing can make to PGWB detection, the right panel of figure~\ref{fig:auto_max_delens_all_exps} shows the signal-to-noise that the considered experiments would yield if they were to reach their maximum delensing potential (meaning that their delensing efficiency is limited by their instrumental resolution and sensitivity rather than by the quality of the lensing potential reconstruction).\\

\subsection{Delensing LiteBIRD with ground-based experiments}
\label{sec:deslensar_litebird_con_ground}

\begin{figure}[tbp]
\begin{floatrow}
\ffigbox{%
\raggedleft
  \includegraphics[width=0.50\textwidth]{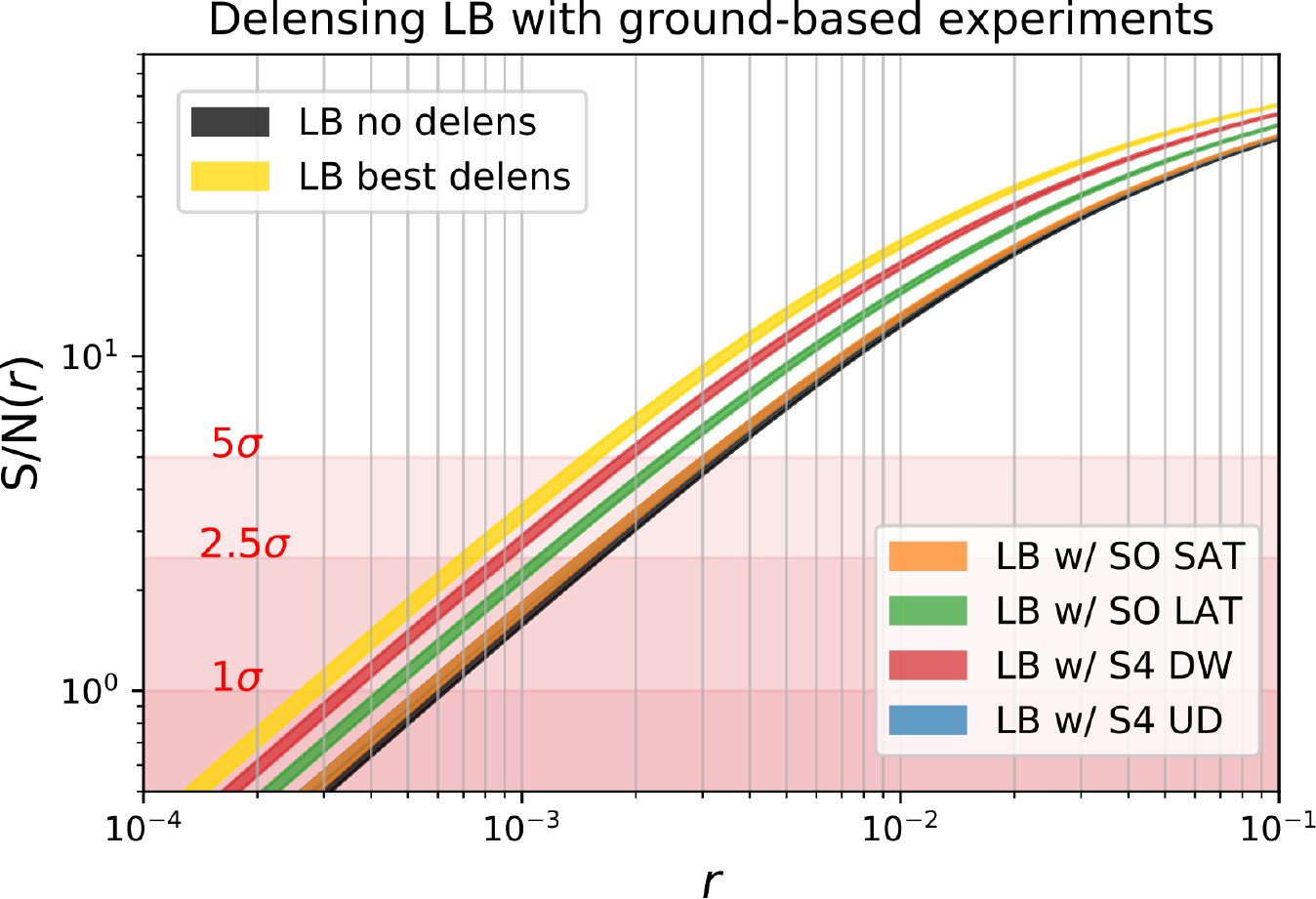}
}{%
  \caption{\label{fig:LB_with_ground} Signal-to-noise of the PGWB detection we could achieve if the LiteBIRD sky was to be delensed with the lensing potential reconstruction coming from ground-based experiments. Contours span the range of signal-to-noise compatible with an $[R_F/3, R_F\times 3]$ uncertainty in the estimation of the foreground residuals. The signal-to-noise LiteBIRD would obtain without delensing, and the one corresponding to LiteBIRD's best delensing (i.e., if $\mathcal{D}_{min}$ was reached) are included for reference. The ``LB w/ SO SAT'' and ``LB w/ S4 UD'' contours are compatible (they overlap) with the signal-to-noise of the non-delensed LiteBIRD sky. }%
}
\capbtabbox{%
    \centering
    \begin{tabular}{|l|c c|}
        \hline
        Experiment & $\mathcal{D}$ & $\mathcal{G}^{0.001}$ \\
        \hline
        SO SAT (@145 GHz) & 0.60 & 1.1 \\
        SO LAT (@145 GHz) & 0.52 & 1.4 \\
        
        S4 DW (@155 GHz) & 0.42  & 1.7 \\
        S4 UD (@155 GHz) & 0.35  & 1.1 \\
        \hline
    \end{tabular}
    \vspace{1.5cm}
}{%
  \caption{\label{tab:delensing_lb_con_ground} Delensing fraction that could be reached if the LiteBIRD sky were to be delensed with an iterative potential reconstruction with the same $S/N_\phi$ as those internally produced from different ground-based experiments. Such delensing would only be possible in the region of the sky where both experiments overlap. $\mathcal{G}^{0.001}$ shows the fractional gain in signal-to-noise that delensing LiteBIRD with these ground-based experiments would report in an $r=0.001$ scenario. 
}}
\end{floatrow}
\end{figure}

Like we saw in the previous subsection, the LiteBIRD satellite does not have a good enough resolution to internally produce a lensing potential reconstruction with which to optimally delens itself (see table~\ref{tab:delensing_forecast}). That does not mean that LiteBIRD cannot be delensed at all (in principle a $\mathcal{D}_{min}=0.33$ could be reached), but rather that an external estimate of the lensing potential is needed for it. Limiting ourselves to potential reconstructions coming from the CMB itself, we could then delens LiteBIRD with the reconstructions coming from the different surveys of the Simons Observatory or the CMB Stage-IV. Table~\ref{tab:delensing_lb_con_ground} collects the delensing fractions that would be reached this way. \\

However, when using external datasets we can only hope to delens the region of the sky where both experiments overlap. Thus, the reduced sky coverage of potential reconstructions from ground-based experiments would make us loose one of the main advantages of space-based observation: the access to the reionization peak (i.e. the $2\leq\ell\leq10$ multipole range). To reduce the impact of this loss of sky coverage, we could still keep the information from the untouched LiteBIRD large scales in the analysis. To do so we will sum the Fisher matrices obtained from considering separately the $\chi^2$ coming from the large scales of the full non-delensed LiteBIRD sky (the $2\leq\ell\leq10$ multipoles of $f_{sky}^{LB}$), the $\chi^2$ from the small scales of the non-delensed region of the LiteBIRD sky (the $10\leq\ell\leq200$ multipoles of $f_{sky}^{LB}-f_{sky}^{ground}$), and the $\chi^2$ from the small scales of the delensed region of the LiteBIRD sky (the $10<\ell\leq200$ multipoles of $f_{sky}^{ground}$). Figure~\ref{fig:LB_with_ground} shows the signal-to-noise for PGWB detection obtained this way.\\

As can be seen in figure~\ref{fig:LB_with_ground}, a good delensing in a small fraction of the sky (like the one the S4 UD survey would provide) will not make a significant contribution to the overall signal-to-noise. It is poorer delensings over larger areas of the sky which makes a substantial boost in signal-to-noise. In this way, the most profitable course of action would be to delens LiteBIRD with data from the S4 DW survey. In fact, a LiteBIRD sky delensed with the internal potential reconstruction coming from the Stage-IV DW survey would have a larger sensitivity to PGWB detection than that of the non-delensed PICO sky. For an $r=0.001$ scenario, this translates into a $\mathcal{G}^{0.001}=1.7$ gain in signal-to-noise with respect to the sensitivity that LiteBIRD could reach on its own without delensing.\\

If we compare the signal-to-noise corresponding to the best delensing possible (if $\mathcal{D}_{min}$ was reached over the whole LiteBIRD sky, yellow contour in figure~\ref{fig:LB_with_ground}) with the signal-to-noise expected for the delensing of LiteBIRD with the different ground-based experiments, we see that there is still room for improvement. Better sensitivities to PGWB detection could be achieved either by enhancing the quality of the potential reconstructions from the ground-based experiments (remember our $S/N_\phi$ are only lower limits to the actual signal-to-noise internal reconstructions could accomplish, and that internal reconstructions can be complemented with other large-scale structure tracers), or by using a potential reconstruction of greater sky coverage (e.g., the CIB could provide a potential reconstruction to cover the full LiteBIRD $f_{sky}$).\\

\subsection{Joint analysis of datasets}
\label{sec:analisis_conjunto}

Instead of using one experiment to delens the other, another possibility to maximize our chances at PGWB detection would be to combine the datasets from the various experiments in a joint analysis. Each experiment will offer a different measurement of the CMB, plagued by the characteristic noise of its instrumental setting and the foreground residuals of the particular component separation technique used. Therefore, we can treat each dataset as an independent measurement of the true CMB and combine them into a composite likelihood. Extending the previous formalism, this is done by adding the Fisher matrices associated to each experiment before obtaining the global $\sigma_r$.\\

\begin{figure}[tbp]
\centering 
\includegraphics[width=1.0\textwidth]{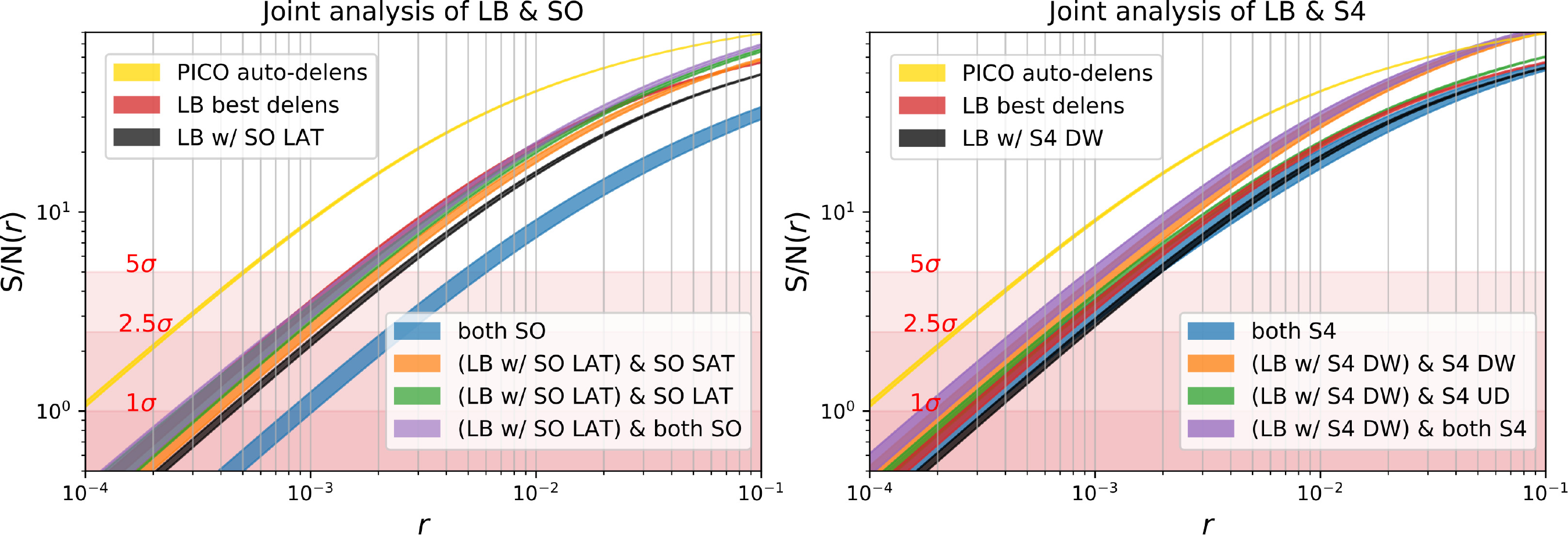}
\hfill
\caption{\label{fig:joint_analysis} Signal-to-noise of the PGWB detection we could achieve through a joint analysis of the data coming from LiteBIRD and the different surveys planned for the Simons Observatory (left) and the CMB Stage-IV (right). Contours span the range of signal-to-noise compatible with an $[R_F/3, R_F\times 3]$ uncertainty in the estimation of the foreground residuals. The ``LB w/ X'' notation indicates when LiteBIRD is delensed with the potential reconstruction coming from experiment X (calculating signal-to-noise like we did in subsection \ref{sec:deslensar_litebird_con_ground}), while ``X \& Y'' refers to the joint analysis of the data from X and Y. With the exception of the SO LAT survey (which cannot be delensed because of its instrumental noise level), ground-based experiments were self-delensed for all the shown combinations. The signal-to-noise PICO would obtain if it were to delens itself is also included for reference.}
\end{figure}

Like we saw in subsection~\ref{sec:cada_experimento_individualmente }, PICO could very well be the ultimate probe for PGWB detection, as long as the tensor-to-scalar ratio falls into the $r\geq2.4\times 10^{-4}$ regime (see figure~\ref{fig:auto_max_delens_all_exps}). However PICO is only a proposal for now, and it would be very interesting to test what values of $r$ could have been already detected through the combination of the other experiments considered here (which are in more advanced phases of development) before it launches. For this purpose, figure~\ref{fig:joint_analysis} shows the signal-to-noise for PGWB detection that the optimal combination of data from LiteBIRD and the Simons Observatory (left panel) and the CMB Stage-IV (right panel) could provide. In both cases, such optimal solution consists of the combination into a composite likelihood of the LiteBIRD sky delensed with the potential reconstruction coming from the ground-based experiment (calculating signal-to-noise like we did in subsection \ref{sec:deslensar_litebird_con_ground}) and the self-delensed sky from the ground-based experiment.\\

As can be seen in the right panel of figure~\ref{fig:joint_analysis}, if they were to use their own internal lensing potential reconstruction to delens themselves, a joint analysis of the data coming from the Ultra-Deep and the Deep-Wide surveys of the CMB Stage-IV would yield a signal-to-noise compatible with that of the best realistic delensing of the LiteBIRD sky (corresponding to the ``LB w/ S4 DW'' contour) in the $r\gtrsim 10^{-3}$ regime. Albeit not shown in figure~\ref{fig:joint_analysis}, if we had combined them without performing delensing first, the signal-to-noise would instead fall slightly below the one corresponding to a non-delensed LiteBIRD sky. This is an example of how delensing can make up for the limited sky coverage of ground-based experiments. \\

Going back to figure~\ref{fig:joint_analysis} and focusing now on the purple contours from both graphs, we can see how the signal-to-noise arising from a joint analysis of ground- and space-based experiments improves upon the one that either of them would achieve separately, reflecting the complementarity of the observations that both settings allow. In this way, a combined analysis of data from LiteBIRD and the CMB Stage-IV could lead to a $2.5\sigma$ detection of an $r=6.0\times 10^{-4}$ (optimal ``(LB w/ S4 DW) \& both S4'' combination in figure~\ref{fig:joint_analysis}). Combining LiteBIRD with both surveys from the Simons Observatory, a $2.5\sigma$ detection of an $r=7.6\times 10^{-4}$ (optimal ``(LB w/ SO LAT) \& both SO'' combination in figure~\ref{fig:joint_analysis}) would also be possible.\\

\section{Discussion and conclusions}
\label{sec:conclusiones}

In preparation for the analysis of the data to come in the next decade, we have compared the performance of the different delensing methodologies and implementations on simulations of the kind of data that the next generation of CMB experiments would provide. With this study we have demonstrated that for next-generation experiments, the delensing efficiency will still be limited by the quality of the data itself rather than by the limitations of current delensing methodologies. Hence, next-generation data could be fully delensed with the already known methodologies and the currently available implementations. Amongst the studied methodologies, we found template delensing within the antilensing approximation to be the optimal technique to employ, since it avoids the computational cost of calculating the inverse displacement without compromising accuracy.\\

Our comparison of lensing/delensing implementations has revealed the systematically better performance that ODE lensing has over the extensively used algorithms based on interpolation in an over-sampled grid. In fact, the only limitations that ODE lensing has shown are those of an implementation constrained to the plane. Therefore, a way to improve the accuracy of the delensing process itself would be to extend the implementation of the ODE lensing algorithm to the full sphere. There is no pressing urgency in the development of such implementation in terms of delensing efficiency, since it is only for extremely good potential reconstructions and very low instrumental noises that an ODE lensing implementation would improve the delensing efficiency of conventional interpolation codes (see the left panel of figure~\ref{fig:comparacion_delensing_fractions}). However, the ODE lensing algorithm has other advantages, such as working on a pixel-by-pixel base, which will facilitate the complex task of working with masked skies, and allowing the exact inversion of the lensing operation by running the ODE in reverse. In addition to the gain on delensing efficiency, delensing via ODE inversion also reduces the computational cost of the delensing process, since delensing is done in a single step instead of the two lensing operations needed for the construction of the B-mode template.\\

We have also studied the dependency that the  delensing efficiency has with the quality of lensing potential reconstructions and the instrumental resolution and sensitivity of the CMB maps that are being used. We have seen that the instrumental beam and sensitivity of CMB experiments constrain the maximum degree of delensing that their data could achieve, and therefore, the selection of a target delensing fraction should be considered as a driver for functional requirements when designing a new CMB experiment. We have also seen that the optimal design for internal lensing potential reconstruction does not necessarily match the optimal design for delensing, the main difference between them being that potential reconstruction requires very fine angular resolutions while delensing does not. This means that experiments with low resolutions would not be able to optimally delens themselves, as it is the case of the LiteBIRD satellite. To overcome this limitation, we could always use the lensing potential reconstructions coming from other CMB experiments or from other large-scale structure tracers to delens our data. In particular, we predict that delensing LiteBIRD with the internal lensing potential reconstruction coming from the Deep-Wide survey of the CMB Stage-IV experiment would report (at least) a $\mathcal{G}^{0.001}=1.7$ fractional gain in the signal-to-noise of the PGWB detection that LiteBIRD could achieve without delensing in an $r=10^{-3}$ scenario. Another important constraint that instrumental specifications impose on delensing efficiency is that, for delensing to successfully reduce the amplitude of lensed B-modes, CMB polarization maps must have an instrumental noise of $\sigma_n\lesssim4\mu K\cdot$arcmin. Above that threshold, delensing will not reduce B-modes below what a simple Wiener filter would already do.\\

Although some works on the literature have already studied some of the aspects related to the role that foregrounds play in the delensing process (e.g., \cite{beck, fabbian, PST}), a complete understanding of the impact of these contaminants on the whole process is still lacking. This is a complex study that is out of the scope of the present paper. In fact, in this work, we have only considered the impact of foregrounds on internal lensing potential reconstructions and on the sensitivity to PGWB detection by including a foreground residual component at the angular power spectrum level (see section~\ref{sec:detectability}). We will study the impact of foregrounds, both, in the internal lensing potential reconstruction, and in the delensing of B-modes, in a future work.\\

It is worth remarking that, not only Galactic foregrounds could degrade the delensing of the CMB anisotropies, but also, extragalactic point sources. We believe that point sources would significantly affect the delensing process for, at least, two reasons. On the one hand, point sources contaminate the smallest scales of the polarization and intensity CMB fluctuations, which are crucial in the reconstruction of the lensing potential. Some preliminary work has already been made to mitigate the effects of point sources in temperature quadratic estimators (see e.g. \cite{PST}). On the other hand, point sources are, by themselves, tracers of the large-scale structure of the universe and, therefore, they could introduce spurious correlations between CMB fields and the gravitational potential. Moreover, these correlations would be enhanced by the actual lensing of the point sources' emission \cite{lensed_foregrounds} (the emission of extragalactic sources at cosmological distances will be lensed by the rest of the matter distribution between them and us). In this sense, methods to detect and characterize point sources in polarization could play a very important role in the delensing of CMB B-modes.\\

\acknowledgments

The authors would like to thank the anonymous referee, whose comments helped to improve the quality of the manuscript. PDP acknowledges partial financial support from the \emph{Formaci\'on del Profesorado Universitario (FPU) programme} of the Spanish Ministerio de Ciencia, Innovaci\'on y Universidades. We acknowledge Santander Supercomputaci\'on support group at the Universidad de Cantabria who provided access to the Altamira Supercomputer at the Instituto de F\'isica de Cantabria (IFCA-CSIC), member of the Red Espa\~nola de Supercomputaci\'on, for performing simulations/analyses. The authors would like to thank the Spanish Agencia Estatal de Investigaci\'on (AEI, MICIU) for the financial support provided under the projects with references ESP2017-83921-C2-1-R and AYA2017-90675-REDC, co-funded with EU FEDER funds, and also acknowledge the funding from Unidad de Excelencia Mar{\'\i}a de Maeztu (MDM-2017-0765).  We make use of the \verb|HEALPix| \cite{healpix},  \verb|CAMB| \cite{camb}, \verb|LensPix|~\cite{lenspix}, \verb|taylens|~\cite{taylens}, \verb|LenseFlow|~\cite{lenseflow} and \verb|quicklens| codes, and the \verb|numpy| and \verb|matplotlib|~\cite{matplotlib} \verb|Python| packages.

\bibliographystyle{JHEP}
\bibliography{main}

\end{document}